\documentclass[12pt,english,preprint,superscriptaddress]{revtex4-1}
\newcommand{\ignore}[1]{} 
\usepackage[latin1]{inputenc}
\usepackage[english]{babel}
\usepackage[T1]{fontenc}
\usepackage[dvips]{graphicx}
\usepackage{color}
\usepackage{subfigure}
\usepackage{ulem}

\date{\today}
\begin{document}

\title{Few-boson tunneling dynamics of strongly correlated binary mixtures in a double-well}

\author{Budhaditya Chatterjee}
\email{bchatter@physnet.uni-hamburg.de}
\affiliation{Zentrum f\"ur Optische Quantentechnologien,  Universit\"at Hamburg, Luruper Chaussee 149, 22761 Hamburg, Germany}

\author{Ioannis Brouzos}
\email{ibrouzos@physnet.uni-hamburg.de}
\affiliation{Zentrum f\"ur Optische Quantentechnologien,  Universit\"at Hamburg, Luruper Chaussee 149, 22761 Hamburg, Germany}

\author{Lushuai Cao}
\email{lcao@physnet.uni-hamburg.de}
\affiliation{Zentrum f\"ur Optische Quantentechnologien,  Universit\"at Hamburg,  Luruper Chaussee 149, 22761 Hamburg, Germany}

\author{Peter Schmelcher}
\email{pschmelc@physnet.uni-hamburg.de}
\affiliation{Zentrum f\"ur Optische Quantentechnologien,  Universit\"at Hamburg, Luruper Chaussee 149, 22761 Hamburg, Germany}

\begin{abstract}
We explore the tunneling dynamics of strongly correlated bosonic mixtures in a one-dimensional double-well. The role of the inter- and intra-species interactions and their interplay  is investigated  using the numerically  exact Multi-Configuration Time Dependent Hartree (MCTDH) method. The dynamics is studied for three initial configurations: complete and partial population imbalance and a  species separated state. Increasing the inter-species interaction leads to a strong increase of the tunneling time period analogous to the quantum self-trapping for condensates. The  intra-species repulsion can suppress or enhance  the tunneling period depending on the strength of the inter-species correlations as well as the initial configuration. Completely correlated tunneling between the two species and within the same species as well as  mechanisms of species separation and counterflow are revealed. These effects are explained by studying the few-body energy spectra as well as the properties of the contributing stationary states. 
\end{abstract}

\maketitle

\section{Introduction}

Ultracold atoms represents a distinct phase of matter for the exploration of fundamental quantum processes \cite{pitaevskii,pethick,bloch07}.  The ability to precisely control ultracold systems has triggered investigations in different fields such as quantum simulation and information processing \cite{buluta09}, quantum phase transitions \cite{greiner02,lewenstein07} and driven quantum systems \cite{kierig08}. Experimentally it is possible to control not only the external potentials  but also the effective interactions between the atoms using Feshbach resonances \cite{chin10}. Moreover,  dimensionality can be tuned and in particular quasi one-dimensional systems can be achieved by confining two transverse degrees of freedom. In such waveguide-like systems, confinement induced resonances  \cite{Olshanii1998a} provide an additional tool to tune the interactions thus making the study of strongly correlated systems experimentally feasible. An interesting observation for a single species bosonic system in one dimension, is that infinitely strongly repulsively interacting bosons possess the same local properties as a system of non-interacting fermions. This effect, known as fermionization has been experimentally observed  \cite{kinoshita04,paredes04} and can be explained via the Bose-Fermi mapping \cite{girardeau60}.

Inspired by the results of single bosonic species, recently, there has been a lot of  experimental \cite{myatt97,hall98,maddaloni00,modugno02,catani08}  and theoretical \cite{cazalilla03, alon06,mishra07,roscilde07,kleine08, girardeau07,zoellner08b,hao08,hao09,tempfli09,mathey09} interest in the static properties of multi-species bosonic mixtures. In these systems, the interplay between the inter- and intra-species forces as well as different masses or potential asymmetry gives rise to various phenomena and effects not accessible in the single component case. For instance, the process of  composite fermionization occurs when the inter-species coupling is set to infinity and the strong repulsion provides  different pathways for phase separation \cite{alon06,mishra07,zoellner08b}. Moreover instabilities \cite{cazalilla03}  as well as new phases such as paired and counterflow superfluidity \cite{mathey09} have been observed.

Focusing on the quantum dynamics, the double well  provides the simplest prototype for a finite lattice and Josephson junction  and is especially  a very elucidating case for studying the fundamental characteristics of  quantum tunneling. Theoretically, the tunneling dynamics of single species through the crossover from weak to strong interaction regimes reveal interesting effects such as  Josephson oscillations, pair tunneling, self trapping as well as fermionization \cite{salgueiro06,dounasfrazer07b,wang08,zoellner07a,zoellner08} which have also been observed experimentally \cite{albiez05,anker05}. 

More recently these studies have been extended also to  systems of binary bosonic mixtures \cite{kuang00,xu08,mazzarella09,satija09,diaz09,sun09,naddeo10,mathey11}. These works demonstrate various effects such as macroscopic quantum self-trapping and coherent quantum tunneling \cite{kuang00}, observations of collapse and revival of population dynamics \cite{sun09,naddeo10}, symmetry breaking and restoring scenarios \cite{satija09} as well as dipole oscillations induced pairing and counterflow superfluidity \cite{mathey11}. 

However, most of the work has been done on the mean-field level either by solving Gross-Pitaevskii equations or by using the lowest band Bose-Hubbard model. Although these studies do provide interesting insights into the mechanism of tunneling, an investigation of the complete crossover from the weak to strong interaction regime  allows the examination of new effects and mechanisms not present e.g. in the mean field description. For instance referring to the case of two species in a harmonic trap, it has been found that if one species is localized due to its heavy mass then it can act as an effective material barrier through which the lighter component tunnels \cite{pflanzer09,pflanzer10}. The feedback of this material barrier leads to different pairing mechanisms for the light species. Moreover, few body systems provide a bottom-up approach towards the understanding of  many-body phenomena. Experiments exploring few atom systems in finite optical lattices \cite{widera11} serve as  promising setups for designing transistor-like structures from the perspective of atomtronics.  

In this paper we study the tunneling dynamics of a binary mixture of bosonic species in a one-dimensional  double-well from a few body perspective. Using the numerically exact Multi-Configuration Time-Dependent Hartree method (MCTDH, see Appendix \ref{sec:mctdh}) \cite{meyer90,beck00}, we investigate the crossover from weak to  strong interactions focusing in particular, on microscopic quantum effects and mechanisms which are prominent in few body systems. We demonstrate how the interplay between the inter- and intra- species interactions affect the rate and behavior of the tunneling in a non-trivial way. A strong increase of the tunneling period is observed as the inter-species repulsion is increased. However in certain cases, and especially for the strongly-interacting regime, increasing the intra-species interactions leads to an increase of the tunneling rates in contrast to what is observed for single species systems. Preparing different initial states leads consequently to a diverse tunneling behavior. For complete imbalance of the populations, i.e., when the particles are all prepared initially in the same well, or when the species are localized at different wells (species separation) the tunneling is strongly correlated meaning that the species tunnel either in phase or out of phase . Only for  very strong intra-species interactions these correlations are reduced. On the other hand for partial population imbalance e.g. one species is delocalized and the second one is localized, a mechanism of species separation and counterflow appears. The various effects are attributed to the features of the energy spectrum and explained by examining the density profile of the contributing stationary states.

The paper is organized as follows. In Section \ref{sec:setup} we introduce our model and setup. Subsequently we present and discuss the results for the quantum dynamics of the mixture with three bosons (two bosons of species A and one of species B). Three initial state scenarios are examined: complete population imbalance  in Sec. \ref{sub:3p_imbalanced}, complete species-separation in Sec. \ref{sub:3p_separated}, and partial imbalance in Sec. \ref{sub:3p_part}. The computational method MCTDH is described in the Appendix \ref{sec:mctdh}.

\section{Model and setup \label{sec:setup}}

We consider a mixture of two species of bosons labeled by $A$ and $B$ in a one-dimensional double well potential. These may correspond to two different kinds of atoms or could be two hyperfine states of the same atomic species. The fact that there are two different species induces distinguishability and thus fundamentally alters the physics and in particular the quantum dynamics  compared to  the case of a  single species.

Our Hamiltonian reads (see \cite{zoellner08b} for details)

\begin{equation}H = \sum_{\sigma=\mathrm{A,B}} \sum_{i=1}^{N_{\sigma}} \left[\frac{p_{\sigma,i}^{2}}{2M_{\sigma}} + U_{\sigma}(x_{\sigma,i}) +  \sum_{i<j}V_{\sigma}(x_{\sigma,i}-x_{\sigma,j}) \right] + \sum_{i=1}^{N_{\mathrm{A}}}\sum_{j=1}^{N_{\mathrm{B}}}V_{\mathrm{AB}}(x_{\mathrm{A},i}-x_{\mathrm{B},j}). \end{equation}\\
where $M_\mathrm{A,B}$ is the mass for species $A$ and $B$, respectively.

We assume here that the different species obey the same single particle  Hamiltonian, i.e., they possess the same mass and experience the same single-particle potential. The double-well potential $U(x) = \frac{1}{2} M \omega x^2 + h\delta_{\omega} (x)$ is modeled as a harmonic potential with a central barrier shaped as a Gaussian  $h\delta_{\omega}(x) = h\frac{e^{-x^2/2 s^2}}{\sqrt{2\pi}s}$ of width $s = 0.5 $ and height $h=8.0$. Dimensionless harmonic-oscillator units i.e., $M_\mathrm{A}=M_\mathrm{B}=1$, $\omega=1$ are employed  throughout. In the ultracold scattering limit, one can approximate the interaction (both intra-$V_{\sigma}$ and inter-species $V_{\mathrm{AB}}$)  with an effective contact potential \cite{Olshanii1998a} 

\begin{eqnarray*}
 V_{\sigma}(x_{\sigma,i}-x_{\sigma,j}) = g_{\sigma}\delta(x_{\sigma,i}-x_{\sigma,j}) \\
V_{\mathrm{AB}}(x_{\mathrm{A},i}-x_{\mathrm{B},j}) = g_{\mathrm{AB}}\delta(x_{\mathrm{A},i}-x_{\mathrm{B},j})
\end{eqnarray*}
Numerically we sample the delta-function as a very narrow Gaussian (choosing of course a spatial grid dense enough to sample this narrow peak). 

The different initial configurations are achieved by adding a tilt to the double-well which can be different for the two species depending on the required state. Thus an individual well could be made  energetically more favorable (tilted) for a certain species. For instance, to prepare a complete imbalance, the double wells of both species are tilted the same way, while to prepare a species-separated scenario $U_A$ is tilted opposite to $U_B$. To prepare the partial population imbalanced state one has to tune the tilt for both species judiciously depending on the given interaction strength such that the required population configuration is achieved. The ground-state is then computed by the relaxation method  and results in the desired initial state. For the study of the dynamics the tilt is instantaneously ramped down to obtain a symmetric double-well at $t=0$.

In order to investigate systematically and in detail the tunneling processes for binary mixtures we consider the simplest non-trivial few-body system consisting of two bosons of species $A$ and one of species $B$. This system captures the most important microscopic quantum dynamical processes occurring for few-body bosonic mixtures.  In this case we have two independent parameters $g_{AB}$ and $g_A$ (since there is only a single boson  B species). When the inter-species interaction $g_{AB}$ is zero, the two components are completely decoupled  meaning that the single B boson will undergo Rabi oscillations between the wells. The A bosons will then follow a correlated two-particle dynamics regulated by the intra-species interaction $g_A$ (This case is not addressed here but has been discussed in detail in the literature \cite{zoellner07a,zoellner08}). Another case which reduces to that of a  single species  is  $g_{AB} \rightarrow g_A$, where the essentials of the tunneling dynamics is  that of three particles of a single species. Our focus is exclusively onto the cases  where we expect significant deviations from the single species scenario.

\section{Complete population imbalance.}
\label{sub:3p_imbalanced}

We begin our study by exploring the quantum  dynamics for an  initial state where  all the atoms are loaded into the left well. As observables, we compute the time evolution of the one-particle density of each species and the resulting population in each well. For the right well we have 

\begin{equation}n_\alpha (t) = N_\alpha {\int_0}^\infty \rho_\alpha(x;t)dx \end{equation}  where $\rho_\alpha$ is the one-body density of the species $\alpha = A,B$ and the total population of the right well is $n_R = n_A + n_B$. Due to symmetry and resonant mechanisms, we  always have a complete transfer of the population of both species between the two wells which happens in most cases according to a periodic pattern with   period $T$. 

%Table \ref{table} summarizes our results for this tunneling period for different values of the parameters $g_A$ and $g_{AB}$, which we discuss and explain in detail next. 

\begin{center}
\begin{table}
  \caption{Tunneling periods for different $g_{AB}$ values for the case of complete imbalance.}
\label{table}
\begin{tabular}{|c|c|c|c|c|}
\hline
\textbf{Period} &  $g_{AB}=0.0$ &  $g_{AB}=0.2$ &  $g_{AB}=5.0$ &  $g_{AB}=25.0$ \\ \hline
$g_A=0.0$ & $2 \times 10^2$ & $1 \times 10^3$ & $9 \times 10^3$  & $1 \times 10^6$ \\ \hline
$g_A=0.2$ & $6 \times 10^2$  & $6 \times 10^3$ & $1 \times 10^4$ & $1 \times 10^6$ \\ \hline
\end{tabular}
\end{table}
\end{center}

\subsubsection{Repulsive interspecies interaction and binding mechanisms}

\begin{figure}

\includegraphics[width=0.55\columnwidth,keepaspectratio]{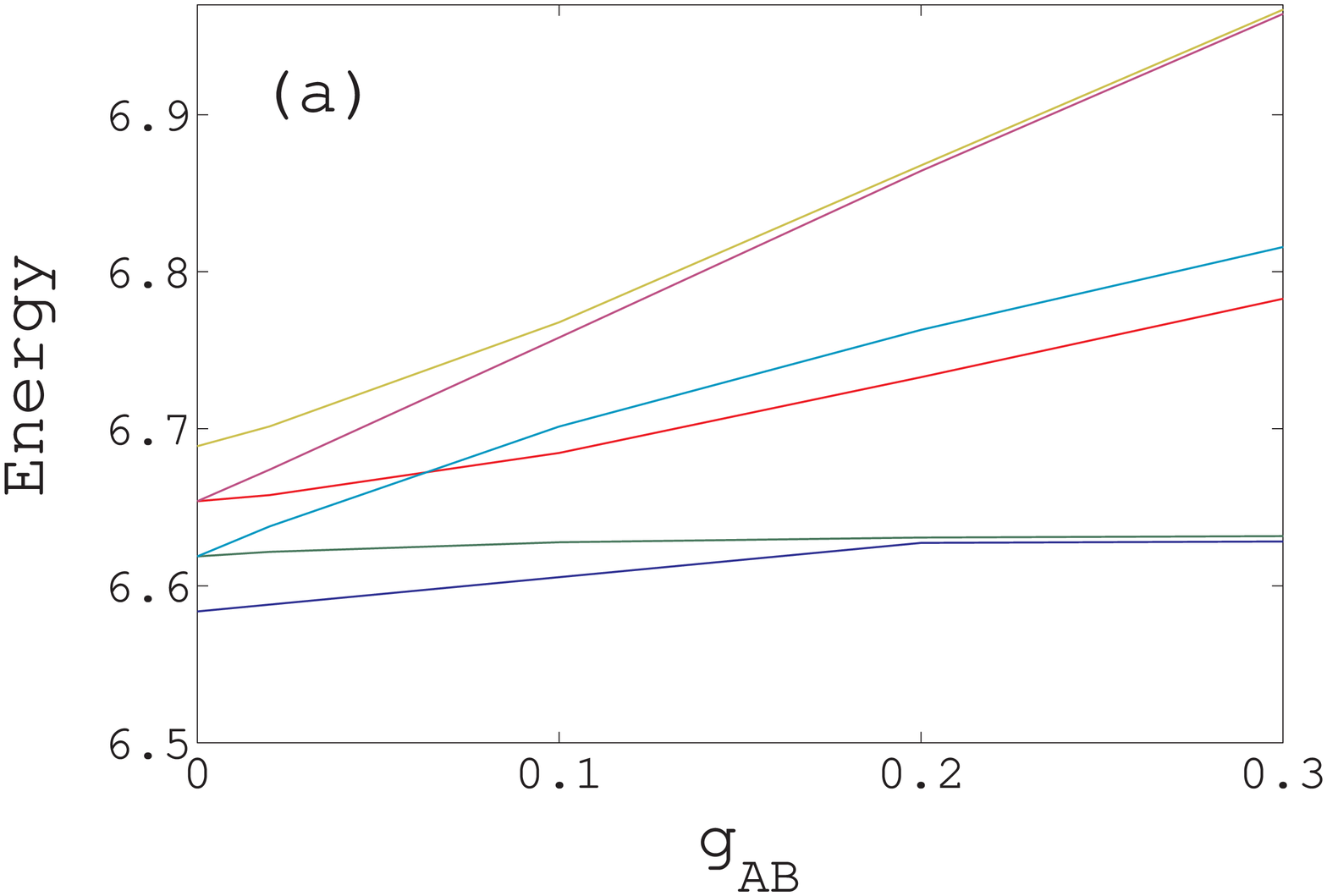}

\caption{(color online) Energy spectrum  for $g_A=0.0$   and  small interaction strengths $g_{AB}$.
\label{cap:3p_energy_ga0_mag}}
\end{figure}

The most important effect of increasing the interspecies interaction $g_{AB}$ is a very strong  increase of the tunneling period up to very large values. This can be seen  in  Table \ref{table}, where we show the tunneling period with increasing $g_{AB}$. This behavior is counter intuitive since with increasing  repulsion between the species initially localized in the same well, one would expect the tunneling to be enhanced. The delayed tunneling is reminiscent of the one found for the case of a single species \cite{zoellner08,zoellner08b} and is the few-body equivalent of self-trapping. The primary reason for this decrease of the tunneling frequency, especially for low interactions (within the so-called Bose-Hubbard regime) can be attributed to the energy spectrum presented in Fig. \ref{cap:3p_energy_ga0_mag} considering the states that contribute to the dynamics. The eigenstates are typically characterized by the superpositions of different number states such as $|AA,B\rangle$, where the vector indicates two A  boson occupying the left well and one B boson in the right well. As $g_{AB}$ increases, different doublets are formed in this lowest band. The energetically highest doublet shown in Fig. \ref{cap:3p_energy_ga0_mag}, consisting primarily of the states $|AAB,0\rangle \pm |0,AAB\rangle$, is of relevance to our case since these eigenstates possess  maximum overlap with our initial state $|AAB,0\rangle$. 

With increasing $g_{AB}$, the number states $|AAB,0\rangle$ and $|0,AAB\rangle$ depart energetically from other number states due to their big on-site interaction energy, (having all bosons in the same well) and   the eigenstates $|AAB,0\rangle \pm |0,AAB\rangle$  become increasingly degenerate thereby forming a doublet. The tunneling then consists  oscillations between $|AAB,0\rangle$ and $|0,AAB\rangle$, while the decreasing energy splitting of the doublet leads to an increase of the tunneling period. This is the few-body analogue of the self trapping mechanism in single-species condensates. The impressive  fact though, is that  this behavior is even more pronounced for higher interactions (see eg. $g_{AB}=5.0,25.0$ in Table \ref{table}), in contrast to the single species case (see ref \cite{zoellner08,zoellner08b}) where there is a reduction of the period due to higher band contributions and fermionization (with the particles tunneling as uncorrelated fermionized bosons with the corresponding Rabi frequency) in the strong interaction limit. In our case, as long as the interaction takes place predominantly between the different species (here for simplicity $g_A=0$), firstly there is no fermionization in the regular sense, and secondly, the particles tunnel in a highly correlated manner - meaning that the initial localized state $|AAB,0\rangle$ does not tunnel to states like $|B,AA\rangle$ or $|AB,A\rangle$ since they possess much lower interaction energy and are thus energetically off resonant. The fact that we encounter correlated particle tunneling (all bosons  together) is documented by the probability of finding all the particles in the same well which remains  very close to unity throughout the dynamics. As a first conclusion, we see that the repulsive inter-species interaction causes 'binding' between the particles and reduces the tunneling rates. 

\subsubsection{Intra-species interactions to control tunneling and correlations}

\begin{figure}

\includegraphics[width=0.42\columnwidth,keepaspectratio]{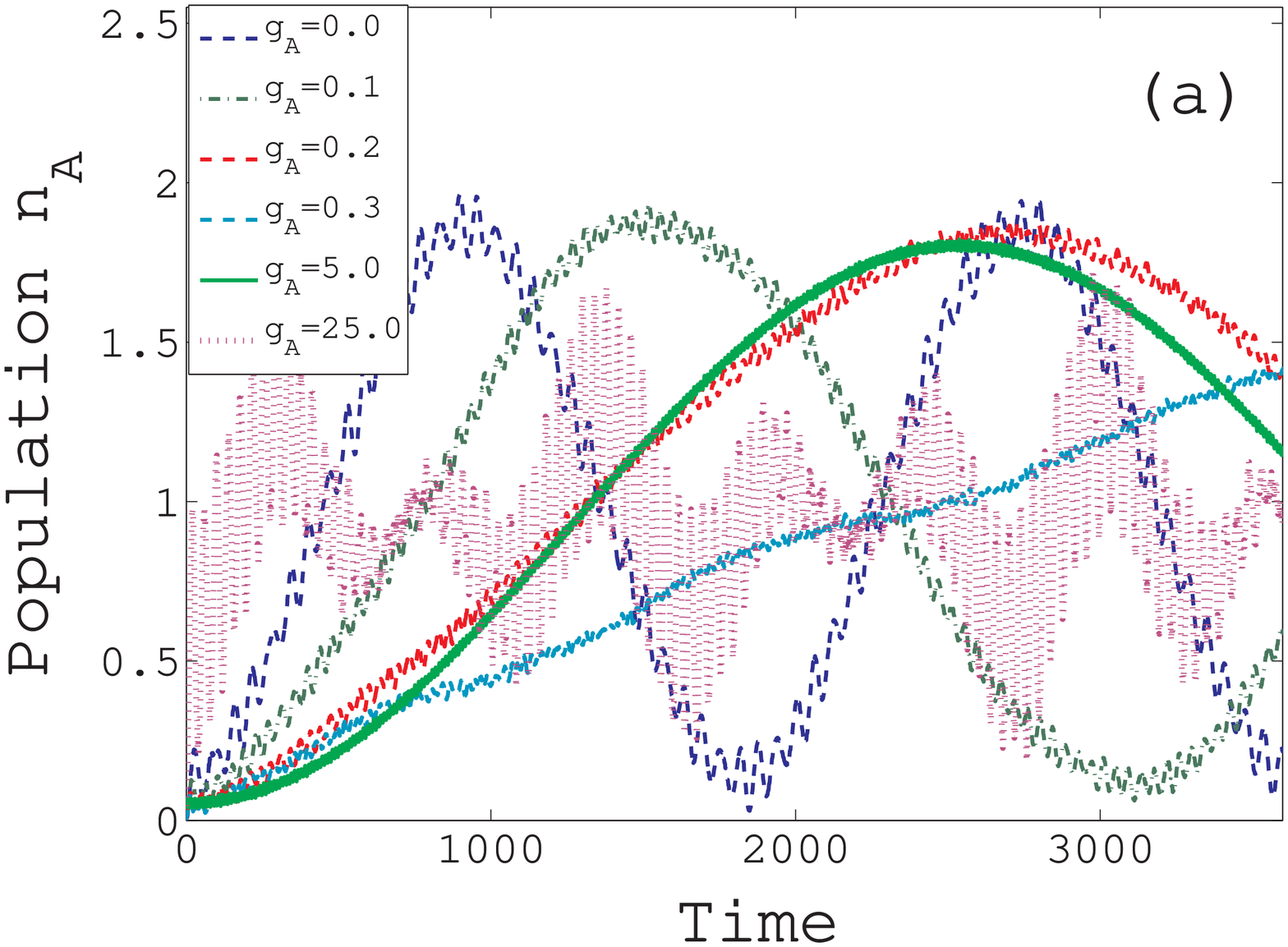}
\includegraphics[width=0.42\columnwidth,keepaspectratio]{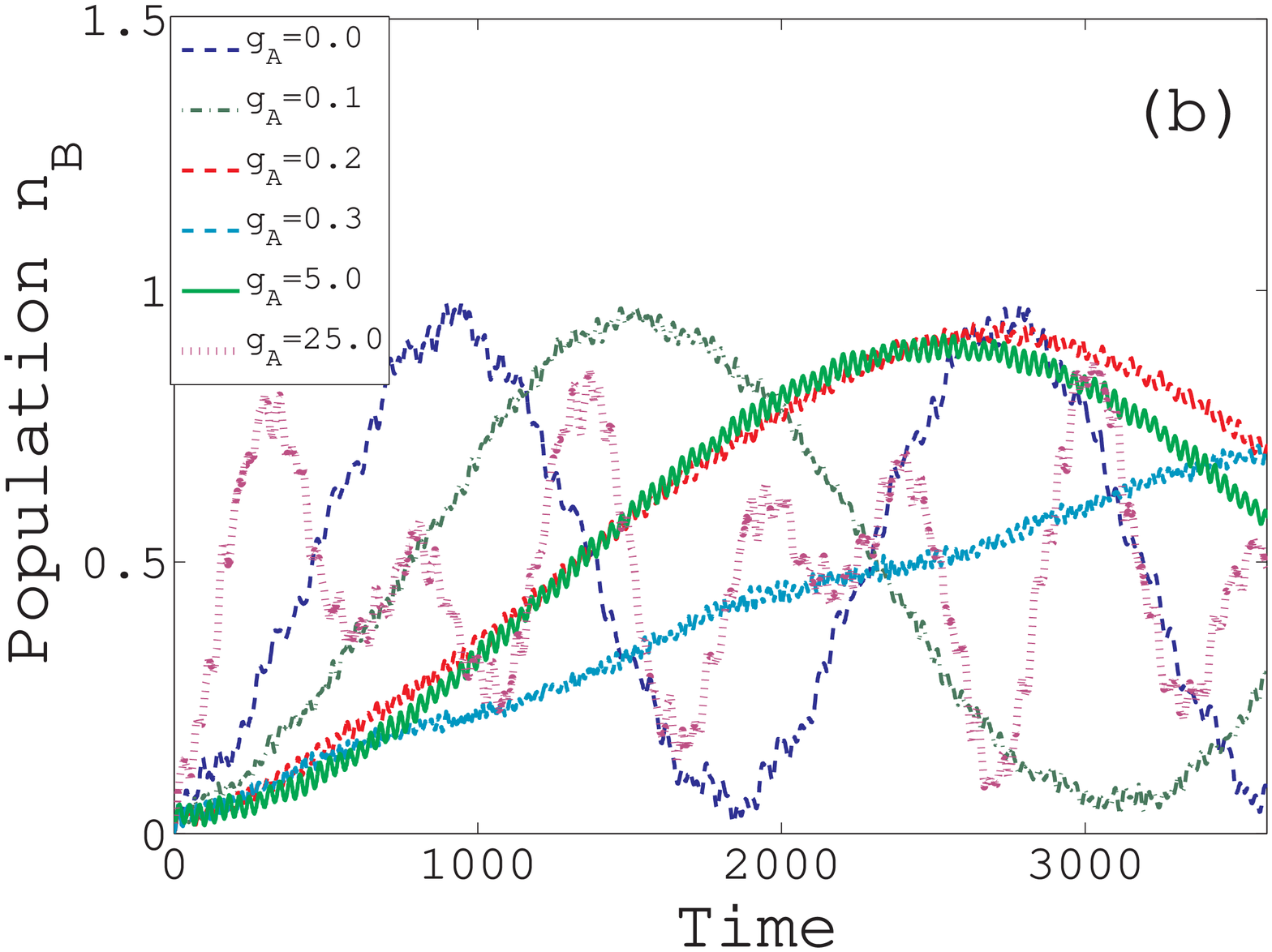}

\caption{(color online)  Population of the right well  (a) $n_A$ of species A and (b) $n_B$ of species B at $g_{AB}=0.2$ for different $g_A$ values. \label{cap:left_3p_gab02}}
\end{figure}

In contrast to the one-way effect of the $g_{AB}$ in the tunneling dynamics, the intra-species repulsion $g_A$ plays a more complicated role - controlling both the tunneling period and the degree of correlations. Let us first explore a weak inter-species interaction strength $g_{AB}=0.2$. In Fig. \ref{cap:left_3p_gab02}, we illustrate  the tunneling dynamics for different values of $g_A$ at $g_{AB}=0.2$ for species A and B by showing  the population of the right well $n_A,n_B$. Except for the periods  $T$ of the tunneling envelope, one also observes  rapid small amplitude oscillations. Concerning $T$, we obtain a monotonic increase as $g_A = 0.0 \rightarrow 0.3$. However this behavior changes as we go beyond the weak interaction regime for $g_A$ and we observe a decrease of the tunneling period for $g_A=5$. Another important feature is that the two components A and B undergo roughly the same evolution of the  oscillation pattern  (compare Fig. \ref{cap:left_3p_gab02} (a) and (b)) which is suggestive of strong inter and intra-species correlations in the sense that all bosons tunnel together. This changes slightly only for very strong interaction $g_A = 25$, where the tunneling period reduces substantially while the pattern becomes more erratic consisting of two primary oscillations and unlike the previous cases, the dynamics of the two components is not completely identical. This indicates, in the line of argumentation provided above, a reduction of the correlations between the two species and attempted single-particle tunneling. 

Within the weak interaction regime where the effective lowest band number states description is valid, the tunneling process of shuffling between the two completely localized states $|AAB,0\rangle$ and $|0,AAB\rangle$ can be more specifically described by the sequence $|AAB,0\rangle \rightarrow |AB,A\rangle \rightarrow |B,AA\rangle \rightarrow |0,AAB\rangle$. The effective tunneling rate  within this lowest band description, is in general approximately given by 
\begin{equation}
\label{effrate}
f \sim J^3/(E_1 -E_2)(E_1 -E_3)\label{eq:tunneling}
\end{equation} 
where $J$ is the effective coupling term between the two sites and $E_1$ is the energy of the initial and final number-state of the tunneling  and $E_2,E_3$ are the energies of the intermediate number-states respectively, which in this case are $|AB,A\rangle$ and $|B,AA\rangle$. Considering the interaction part of the number states, the completely localized states  $|AAB,0\rangle$ and $|0,AAB\rangle$ have energies $\sim$ $2g_{AB}+g_A$ while the states $|AB,A\rangle$ and $|B,AA\rangle$ have energies $\sim$ $g_{AB}$ and $g_A$ respectively. Therefore, for this tunneling process, the tunneling rate according to Eq. (\ref{effrate}) scales as $f\sim J^3/2g_{AB}(g_{AB}+g_A)$. From this relation follows that the tunneling rate decreases for increasing $g_A$ in the weak interaction regime as we have seen above. 

\begin{figure}[htb]

\includegraphics[width=0.45\columnwidth,keepaspectratio]{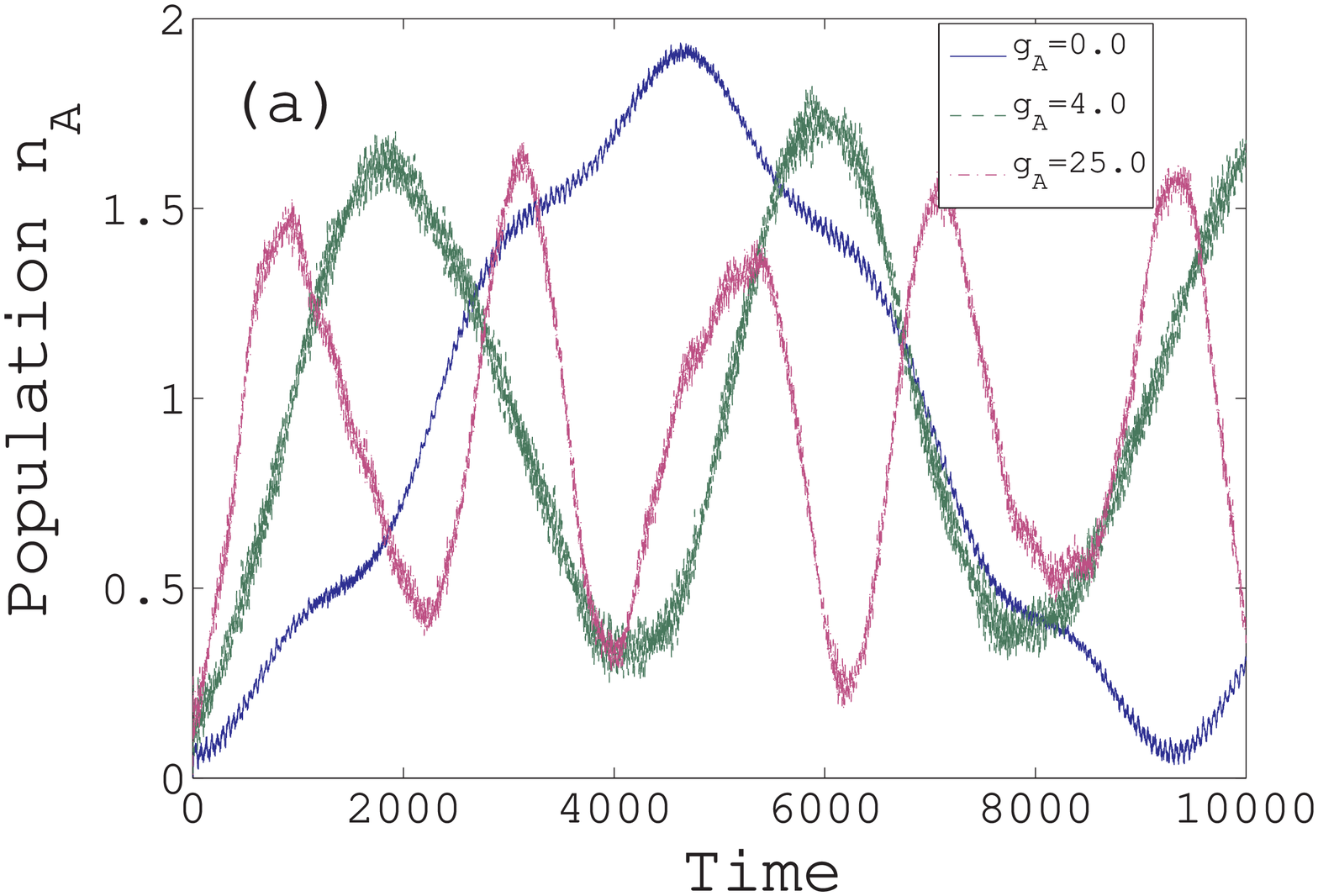}
\includegraphics[width=0.45\columnwidth,keepaspectratio]{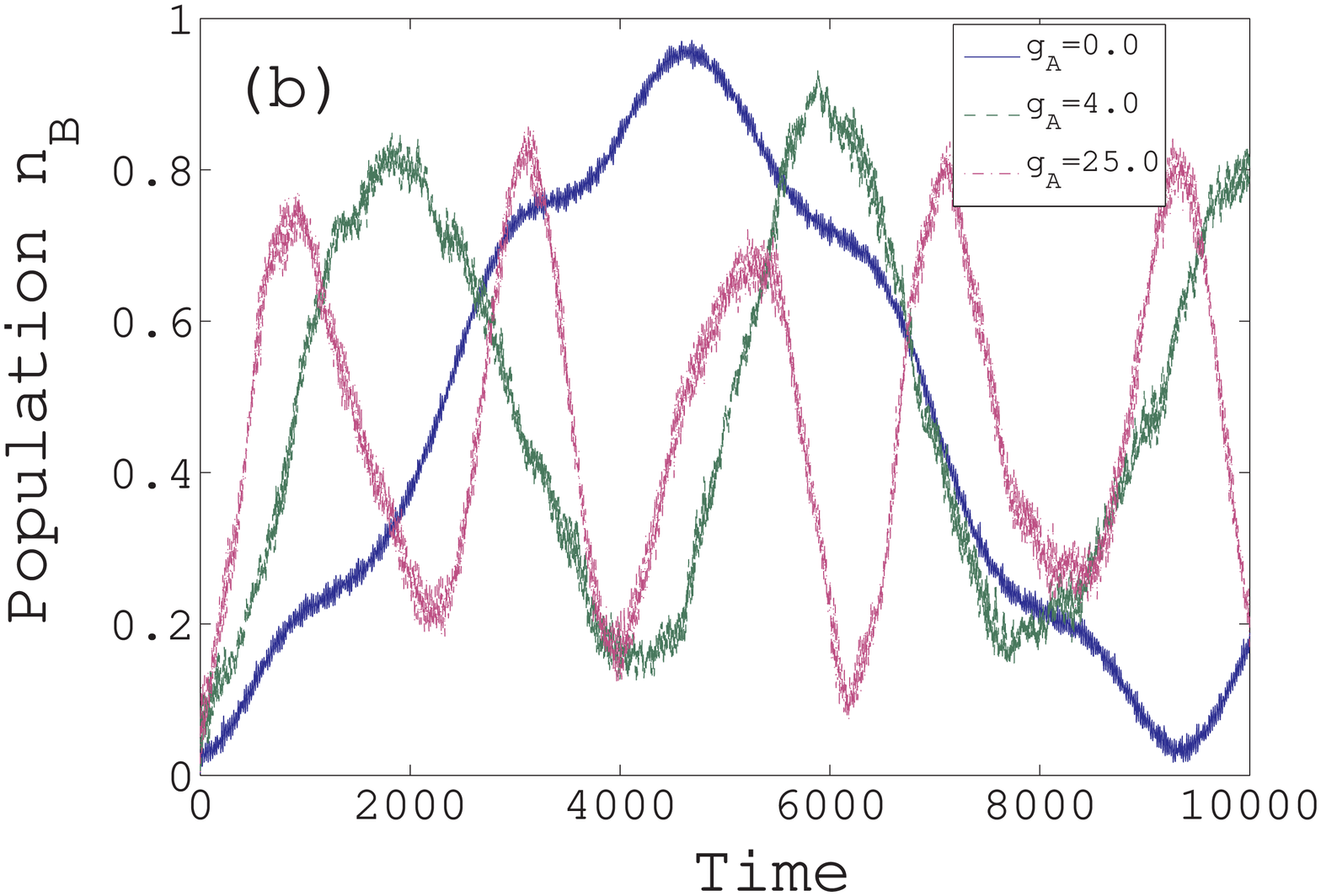}

\caption{(color online) Population in the right well  (a) $n_A$ of species A and (b) $n_B$ of species B at $g_{AB}=5.0$ for different $g_A$ values. \label{cap:3p_gab5}}
\end{figure}

The opposite effect i.e. decrease of the tunneling period, as $g_A$ increases further beyond the weak coupling regime, is attributed to the increasing splitting of the main contributing doublet  $|AAB,0\rangle \pm |0,AAB\rangle$  as $g_A$ increases. Additionally, higher bands contributions appear, especially for stronger intra-species  interactions $g_A=25.0$, which breaks the completely correlated tunneling behavior, allowing for attempted single particle tunneling into energetically higher number-states like  $|AB,A\rangle$. 
\begin{figure}

\includegraphics[width=0.45\columnwidth,keepaspectratio]{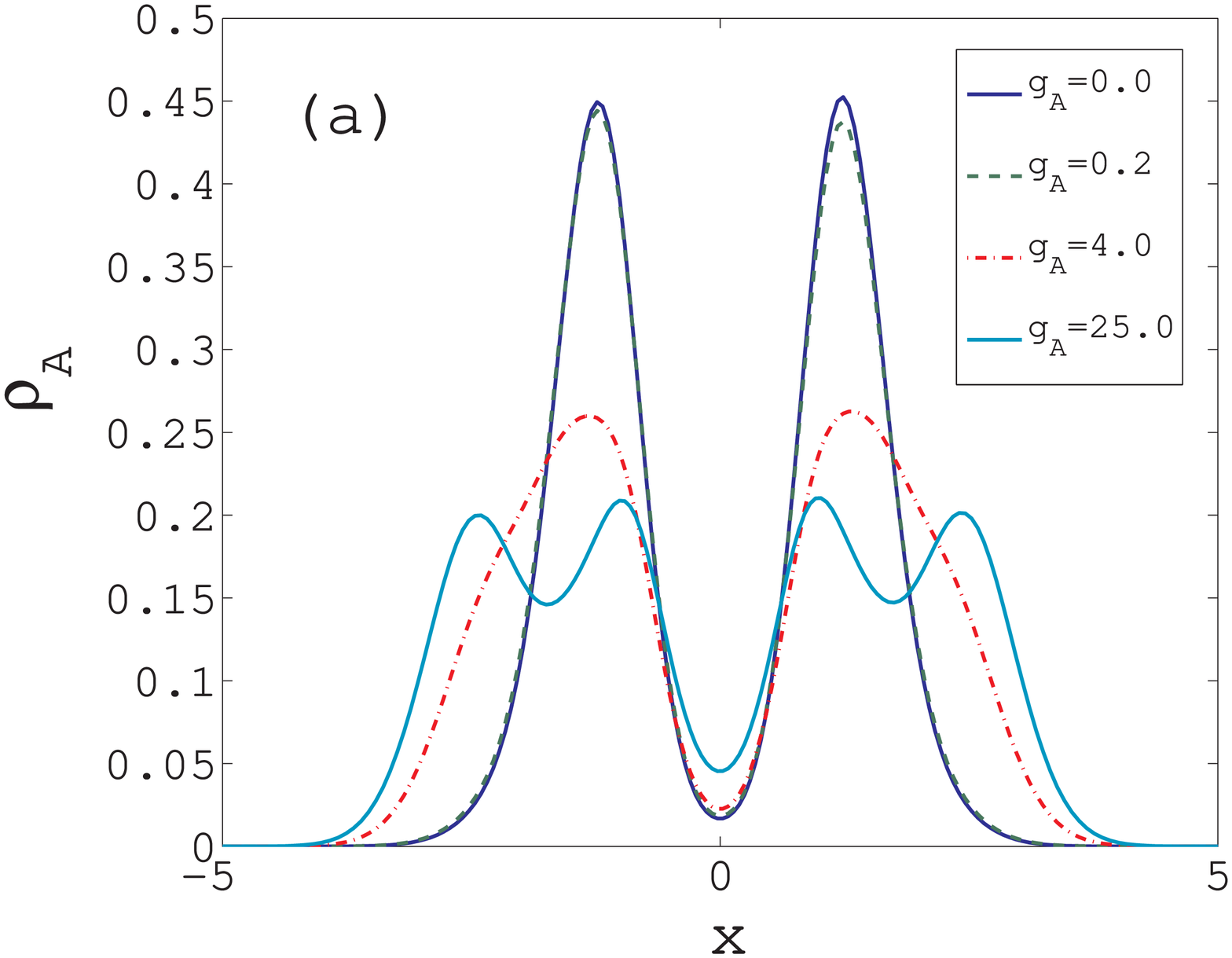}
\includegraphics[width=0.45\columnwidth,keepaspectratio]{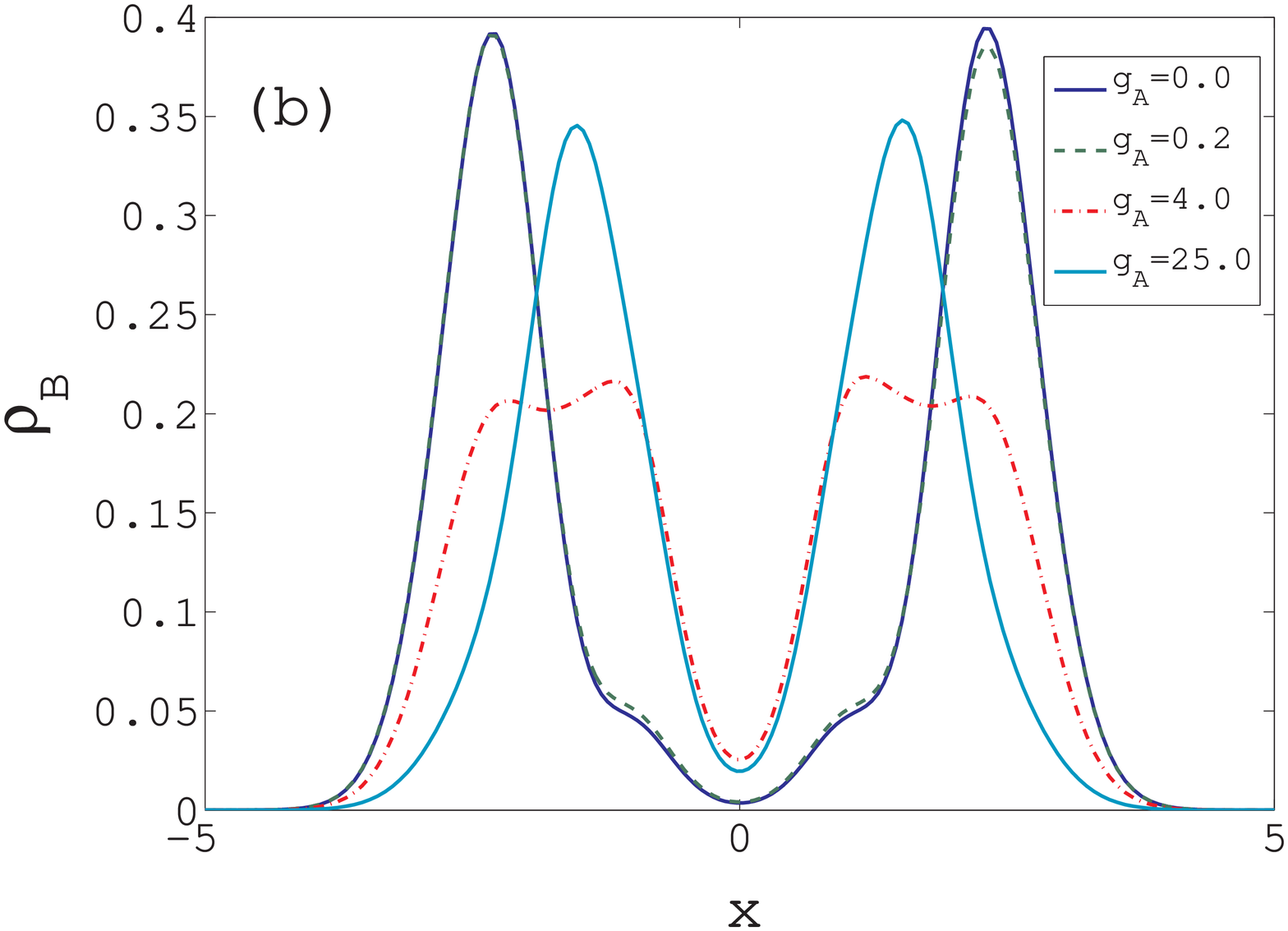}

\caption{(color online)  One-particle density as obtained by taking into account the most important eigenstates contributing to the initial state for (a) species A  and (b) species B  for $g_{AB}=5.0$.
\label{cap:1p_density_left}}
\end{figure}

Turning now to higher inter-species interaction $g_{AB}=5.0$  we observe in Fig. \ref{cap:3p_gab5}  that the tunneling period decreases strongly as  $g_A$ increases. Since $g_{AB}$ is in this case beyond the weak coupling regime, we focus on an analysis of the density profiles of the contributing eigenstates shown in Fig. \ref{cap:1p_density_left}  to understand the effect of increasing $g_A$. As $g_A$ increases, the repulsion of the A bosons leads to a broadening of their density profile. This broadening leads to a greater overlap of the  wave functions of A atoms  localized in the left and the right well and this in turn increases the effective tunneling coupling and the corresponding tunneling rates. At $g_A=0$, the localized densities $\rho_A$ and $\rho_B$ are spatially separated in each well as a consequence of the repulsion between the species. Note that the density of the  $B$ boson possesses its maximum for larger values of $|x|$ thereby 'sandwiching' the $A$ boson population. This arises from the fact that due to the unequal number ($N_A>N_B$), it is energetically favorable to shift the density of the $B$ species to  larger values of $|x|$. As $g_A$ is increased,  the two localized densities $\rho_A,\rho_B$ in the two wells gain an increasing overlap which can be observed as a vertical upward shift of the density profile at $x=0$ that becomes progressively stronger with increasing $g_A$. This mechanism, also present for other contributing states, leads to an overall increase of the tunneling coupling and consequently to an increase of the tunneling frequency for strong interactions. 

\begin{figure}

\includegraphics[width=0.45\columnwidth,keepaspectratio]{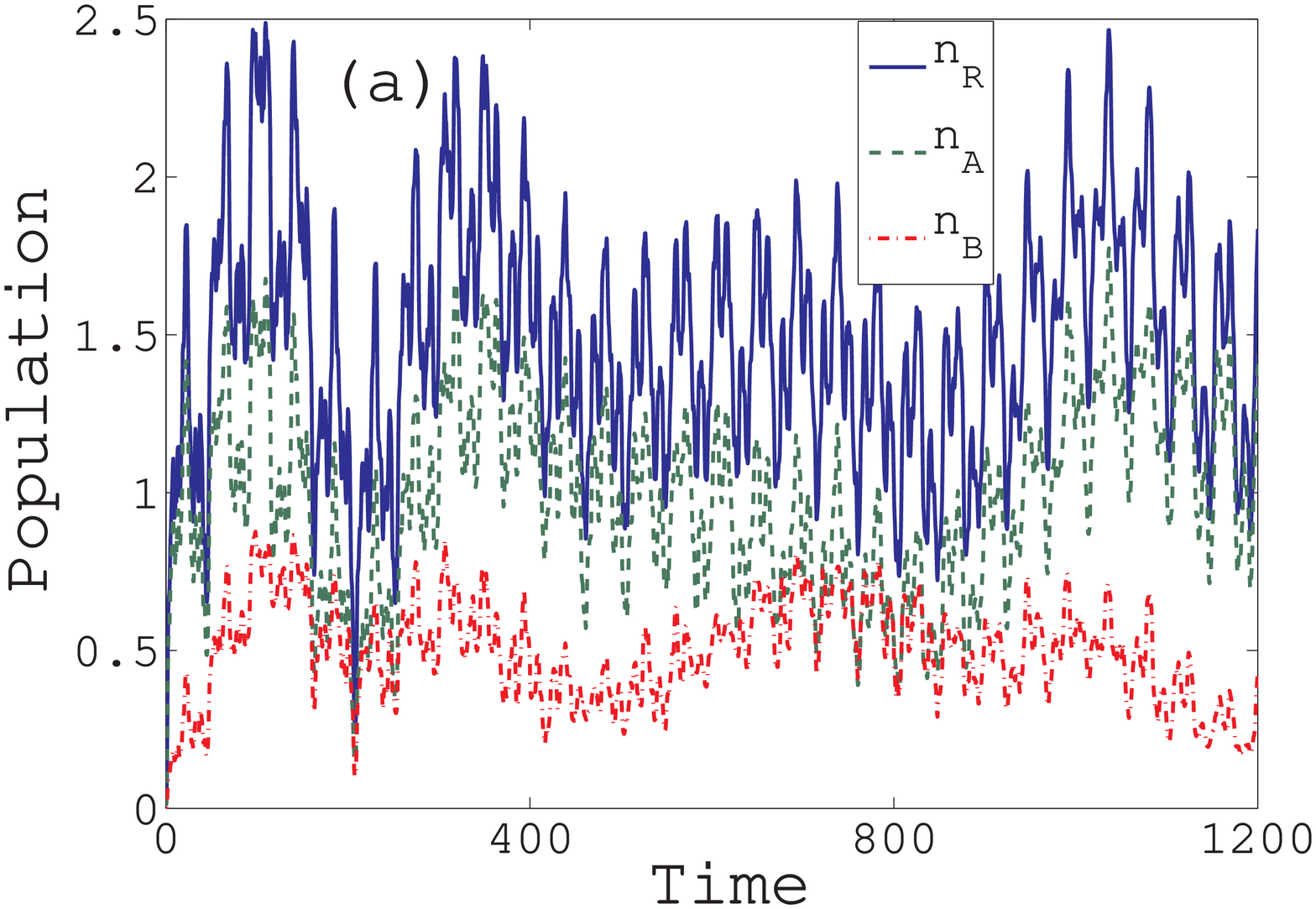}
\includegraphics[width=0.47\columnwidth,keepaspectratio]{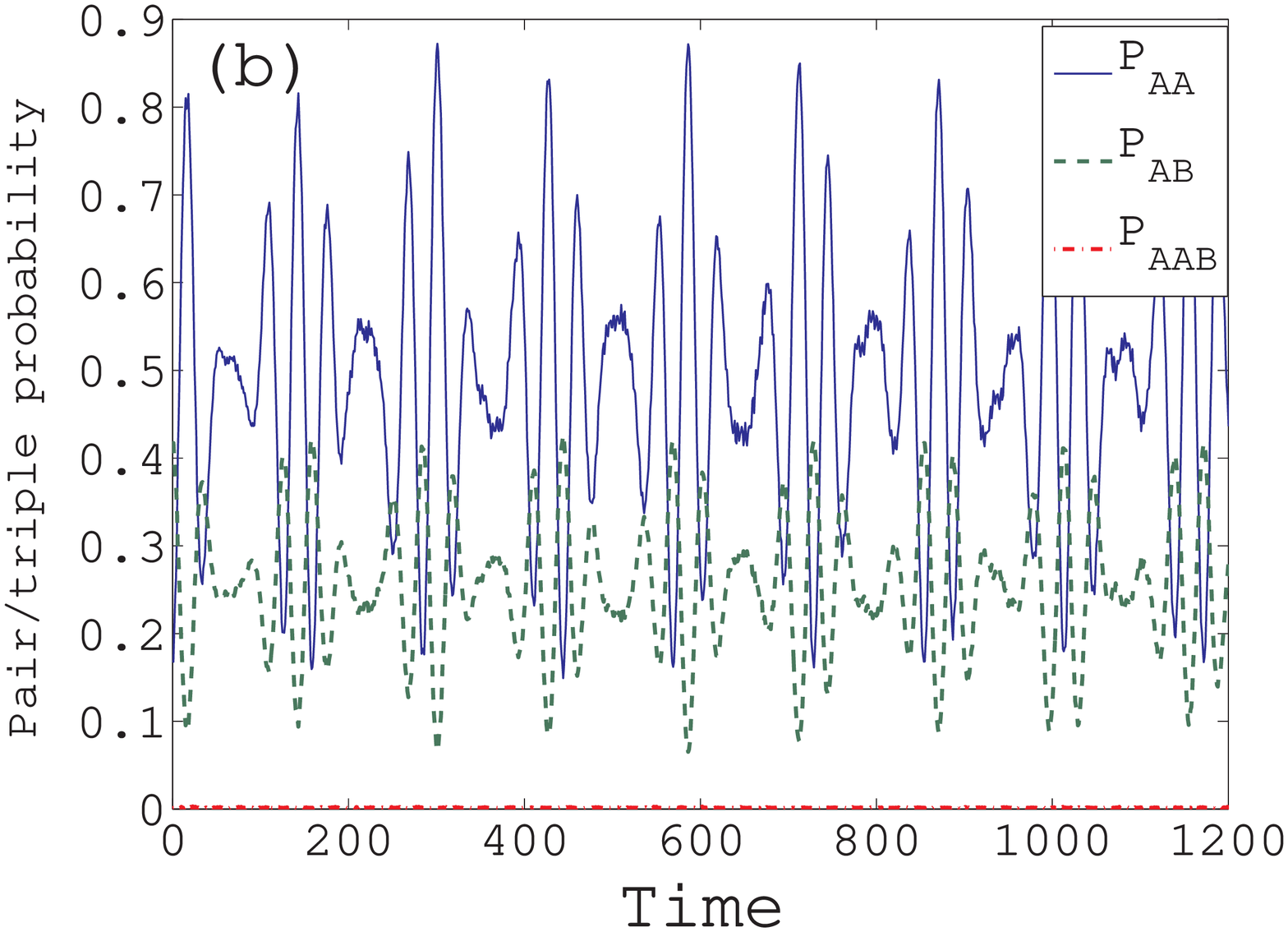}

\caption{(color online) (a) Population in the right well of the total $n_R$ and the individual species $n_A$, $n_B$ for $g_{AB}=25.0$  $g_A=20.0$. (b) Pair/triple probability. $P_{AA}$, $P_{AB}$ and $P_{AAB}$ correspond to the probability of finding AA, AB and AAB boson in the same well, respectively.
\label{cap:3p_gab25ga20_left}}
\end{figure}

The overall features with respect to the different time-scales and oscillatory tunneling behavior is similar for very strong interspecies interactions $g_{AB}=25.0$ with the exception of  $g_A=20.0$ [Fig.\ref{cap:3p_gab25ga20_left}(a)]. Only this case can be considered as a tunneling mechanism close to fermionization.  In this regime,  the bosons become isomorphic with non-interacting fermions and thus the tunneling dynamics approaches that of independent non-interacting fermions. The latter tunneling frequency is close to the Rabi-frequency which is significantly faster than the previously discussed cases. To understand the reduction of period from a number-state perspective we note that the nearly fermionized bosons occupy both the lowest band and the first excited band. As a consequence, the previously off-resonant intermediate number-states (namely the states $|AAB,0\rangle$, $|AB,A\rangle$ and $|B,AA\rangle$) become near resonant, since the particles can tunnel between the excited band of the two wells without significant change of energy. This results in a reduction of the effective tunneling period while the separation of the time-scales involved in the dynamics is strongly reduced. The correlations among the bosons with respect to the tunneling process is also strongly reduced as can be seen in  Fig. \ref{cap:3p_gab25ga20_left}(b), where the pair and triple correlations show a strong deviation from  the value 1, i.e. from the strongly correlated case.

\begin{figure}
\includegraphics[width=0.45\columnwidth,keepaspectratio]{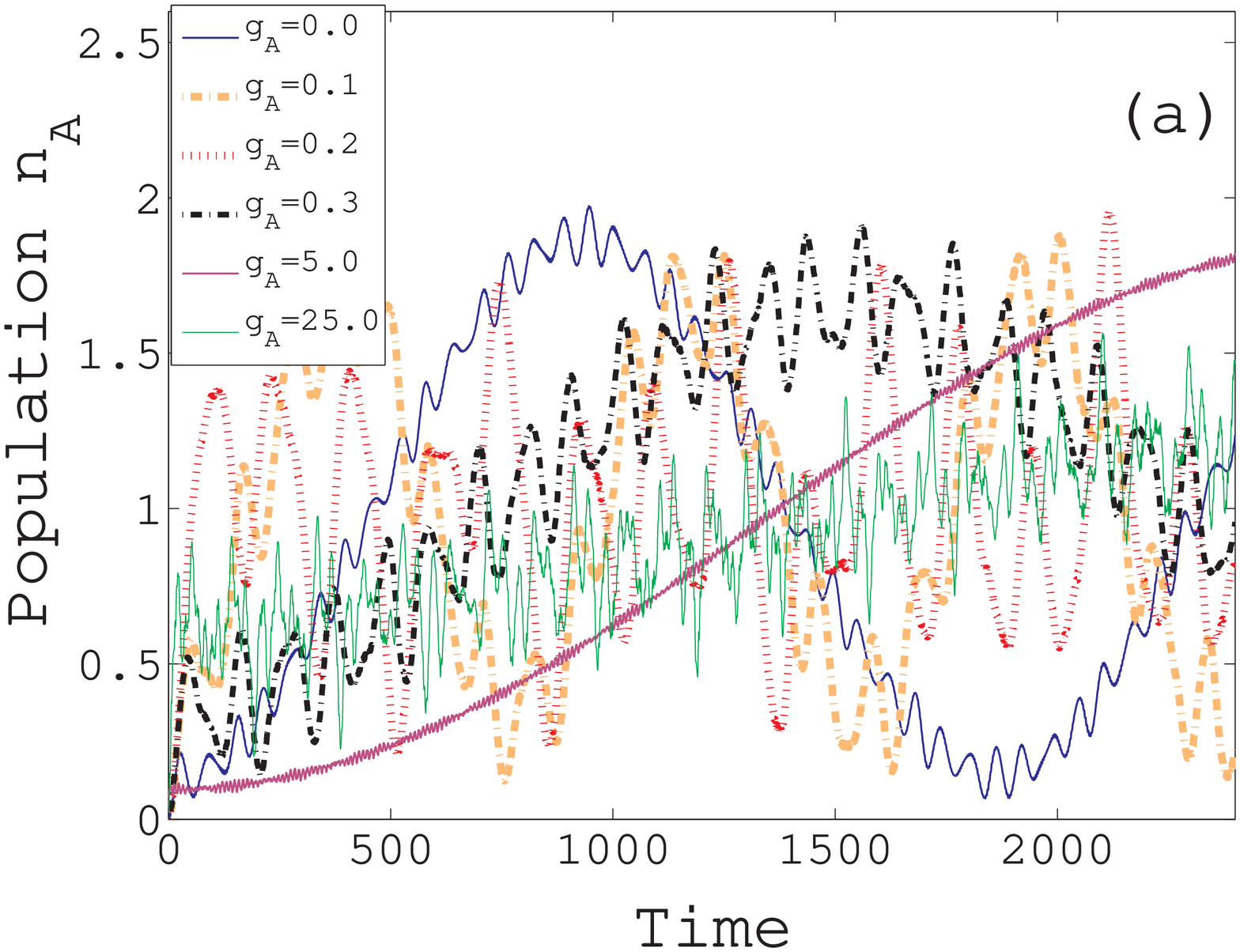}
\includegraphics[width=0.45\columnwidth,keepaspectratio]{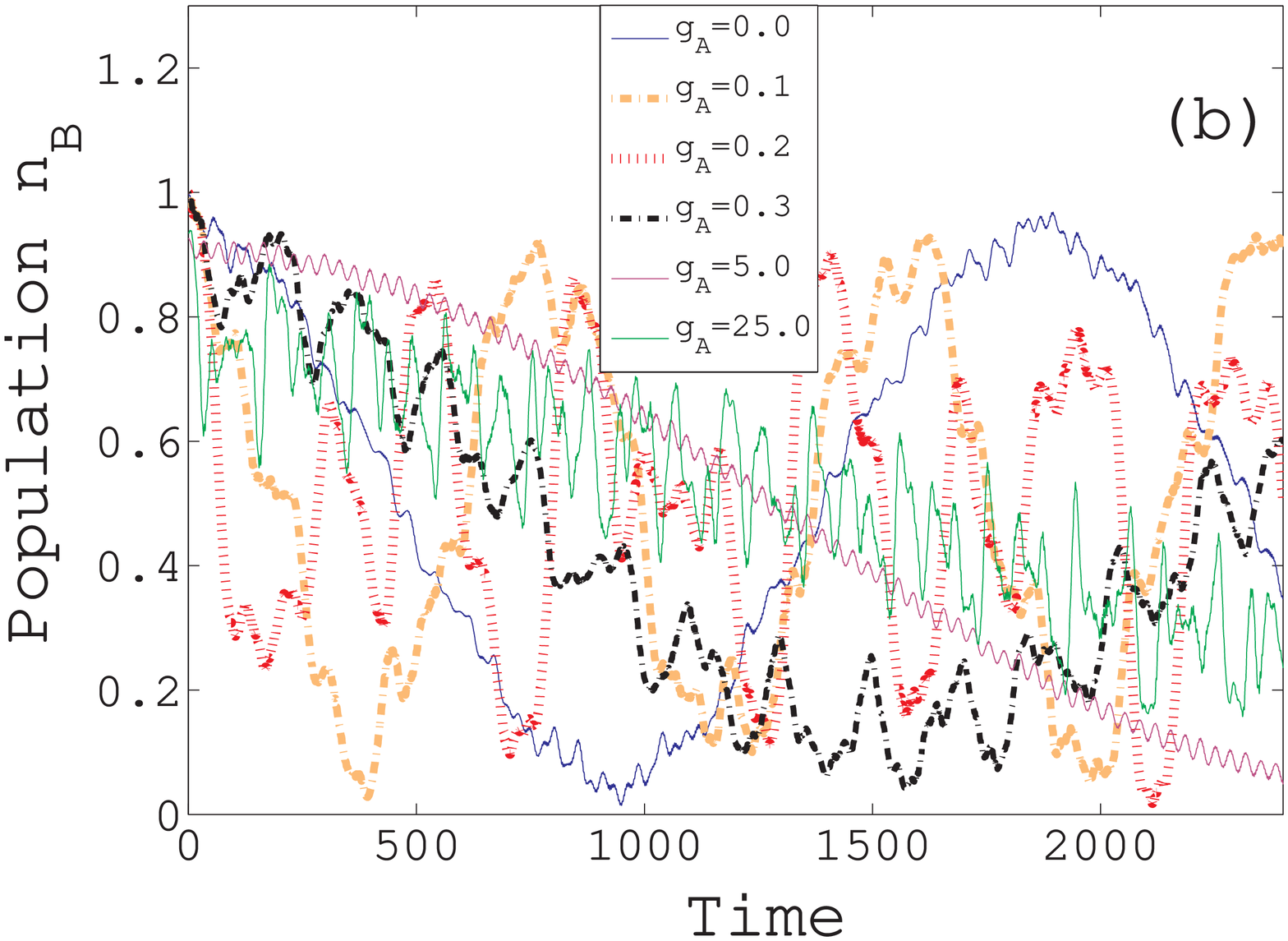}
\includegraphics[width=0.55\columnwidth,keepaspectratio]{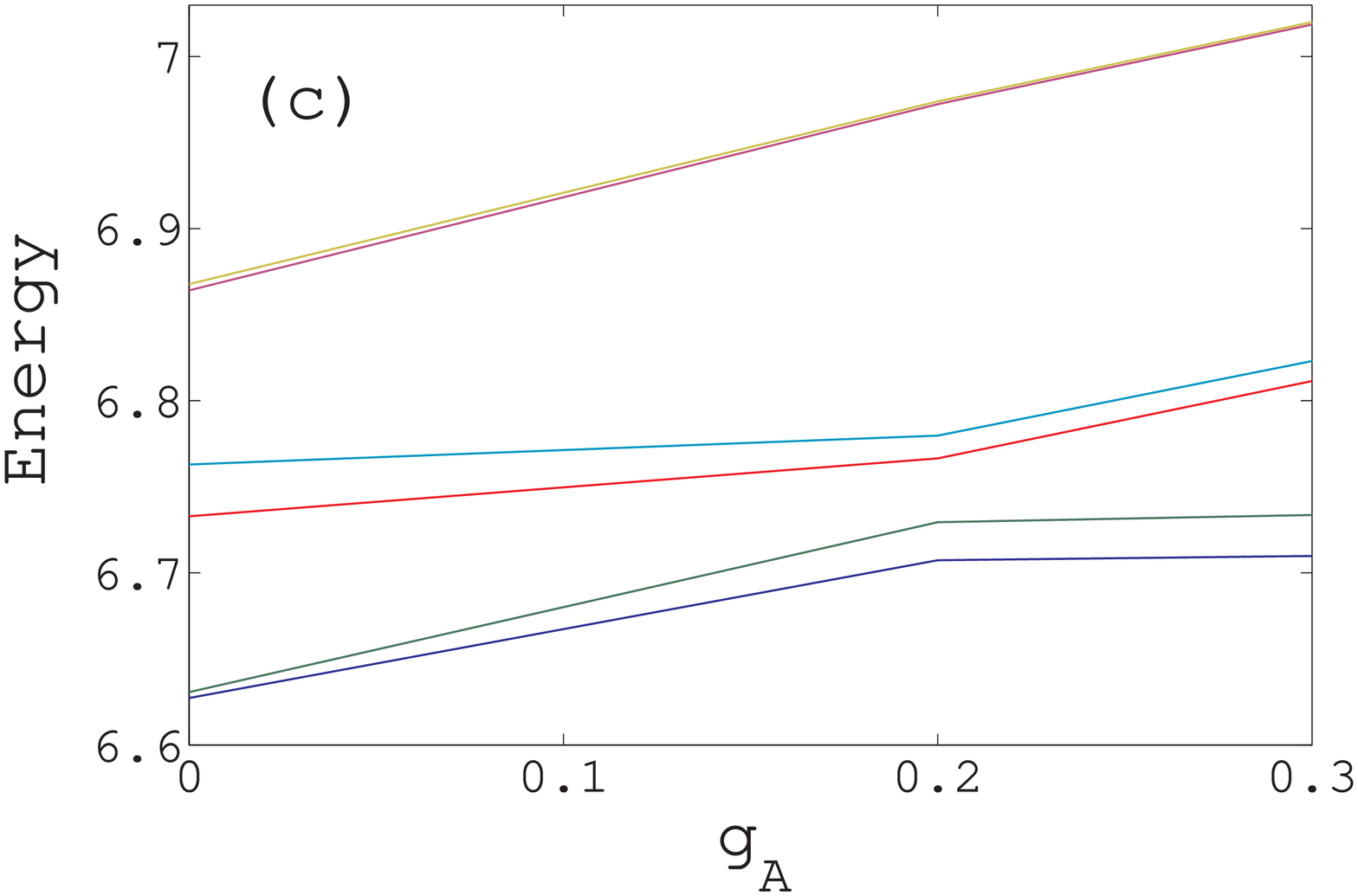}
\caption{(color online)  Population of the right well (a) $n_A$ of species A and (b) $n_B$ of species B at $g_{AB}=0.2$ for different values of $g_A$ for a species separated initial state (c) Energy spectrum for $g_{AB}=0.2$  \label{cap:3p_gab02_sleft}}
\end{figure}

\section{Species-separated initial state.}
\label{sub:3p_separated}

Let us now consider the initial state for which the two species are localized in different wells the $A$ bosons in the left and the $B$ boson in the right well. Similar to the previous scenario, increasing $g_{AB}$ leads again to an increase of the tunneling period, an effect which is intuitive here since the components that are initially prepared in different wells, are forced to stay apart from each other by the repulsive interspecies force. Moreover, an important point to note  is that the contributing states are always those of the lowest band and the number states mainly involved are  $|AA,B\rangle$ and $|B,AA\rangle$  since the former one is the initial state. Therefore as long as  $g_A$  remains comparatively small, the dynamics consists of a  slow oscillation  between the states $|AA,B\rangle$ and $|B,AA\rangle$ and is correlated in the sense that the $A$ and $B$ bosons always occupy different wells in the course of the dynamics. 

The above is shown in Fig. \ref{cap:3p_gab02_sleft} (a),(b) for  $g_{AB}=0.2$, where the population of $A$ and $B$ bosons in the right well is plotted. An important difference compared to the completely imbalanced preparation, is that the increase of $g_A $ leads here to a decrease of the tunneling period $T$ initially, reaching a minimum at $g_A \approx0.2$. Subsequent increase of $g_A$ leads to an increase of the period again.  Resorting to the energy spectrum for an explanation  (Fig. \ref{cap:3p_gab02_sleft}(c)), one should focus on the lowest doublets which have dominant contributions from $|AA,B\rangle$ and $|B,AA\rangle$. We see a splitting of the lowest doublet as they approach the avoided crossing leading to an increase of the tunneling rates. For larger $g_A$, it is the energetically excited doublets which represent the main contribution. The two levels of the excited doublet come closer in energy as $g_A$  increases further leading to a smaller tunneling frequency. In terms of tunneling processes, the dominant sequence here is $|AA,B\rangle \rightarrow |A,AB\rangle \rightarrow |AB,A\rangle \rightarrow |B,AA\rangle$ and a somewhat suppressed sequence is $|AA,B\rangle \rightarrow |A,AB\rangle \rightarrow |0,AAB\rangle \rightarrow |B,AA\rangle$. Using Eq. (\ref{effrate}), the tunneling rates scales as $f\sim J^3/2g_{AB}(g_{AB}-g_A)$ for the first sequence and $f\sim J^3/(g_{AB}-g_A)^2$ for the second sequence. We see that, as $g_A$ increases from zero, the tunneling frequency increases, reaching a maximum for  $g_{AB}=g_A$ (it does not actually diverge as the formula suggest since in this case other higher order terms with respect to $J$ becomes relevant) while beyond this point the period increases again. The crucial difference to the previous case of a completely imbalanced initial condition is the sign in the denominator which was positive previously and is negative here. This implies that while in previous case, there was a monotonic increase in the tunneling period with increasing $g_A$, here we encounter an initial decrease as $g_A$ approaches $g_{AB}$. For very high values $g_A =25.0$  additional states contribute to the dynamics leading to the high frequency 'noise' observed. 

\begin{figure}
\includegraphics[width=0.45\columnwidth,keepaspectratio]{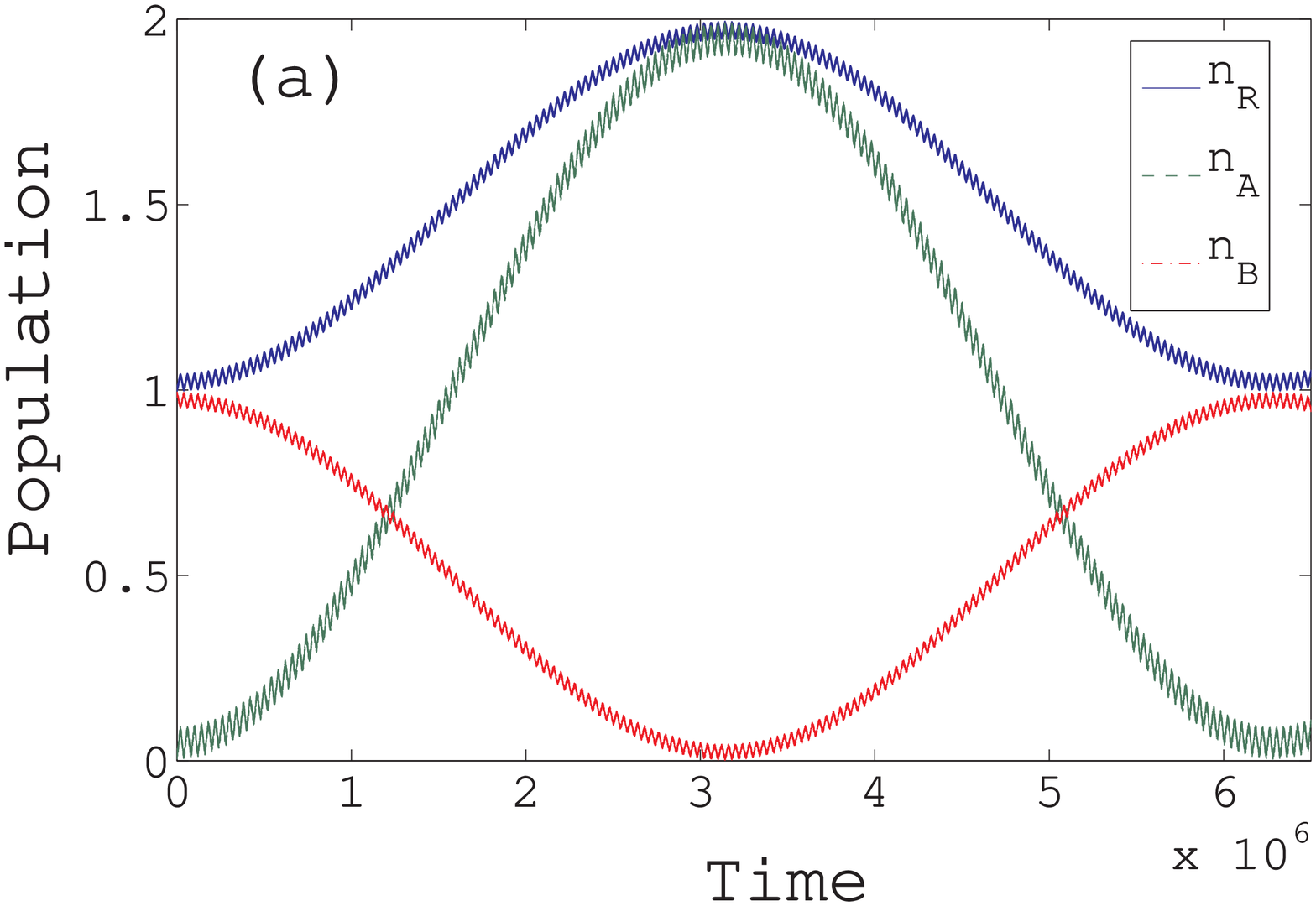}
\includegraphics[width=0.45\columnwidth,keepaspectratio]{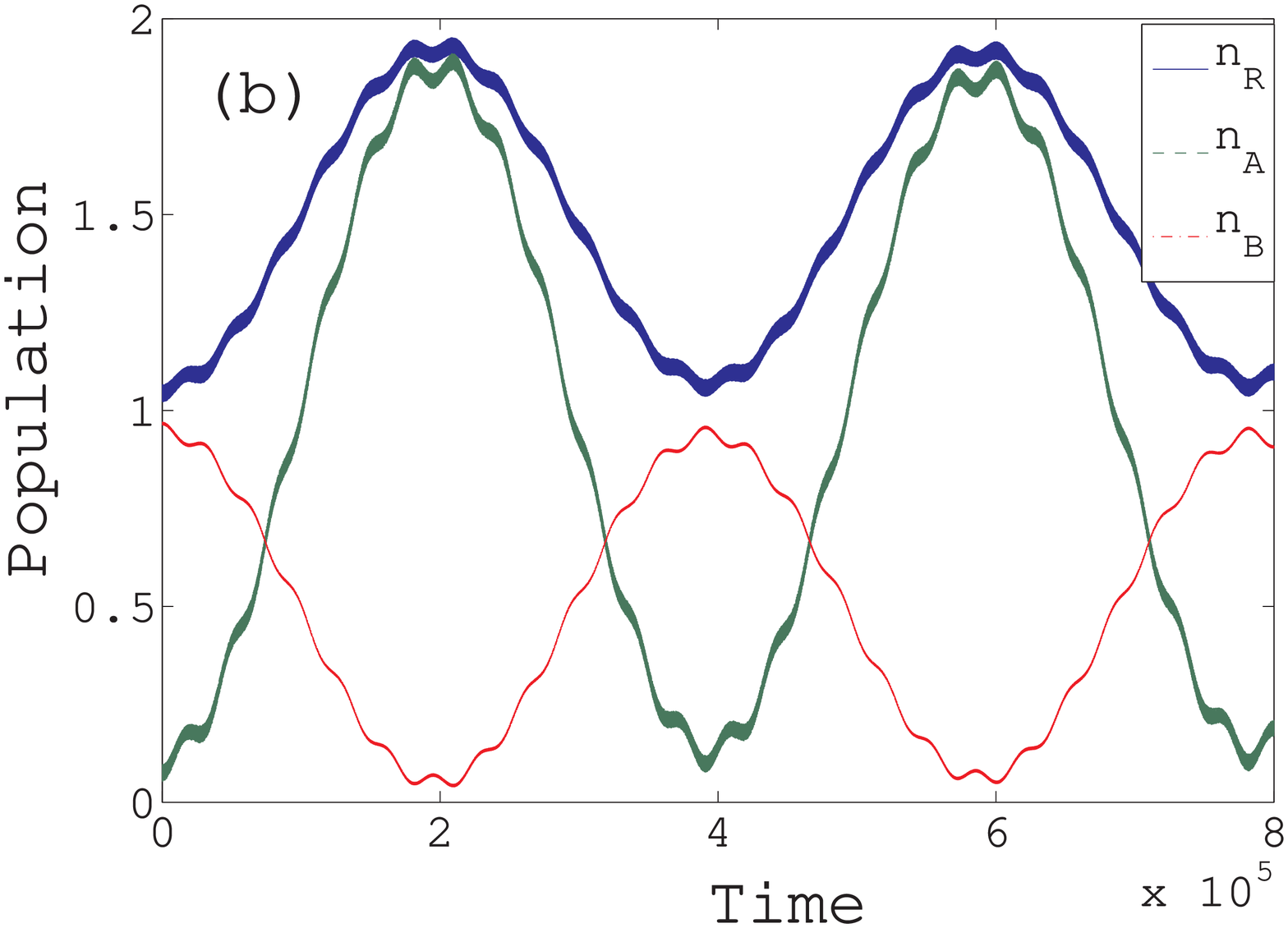}
\includegraphics[width=0.45\columnwidth,keepaspectratio]{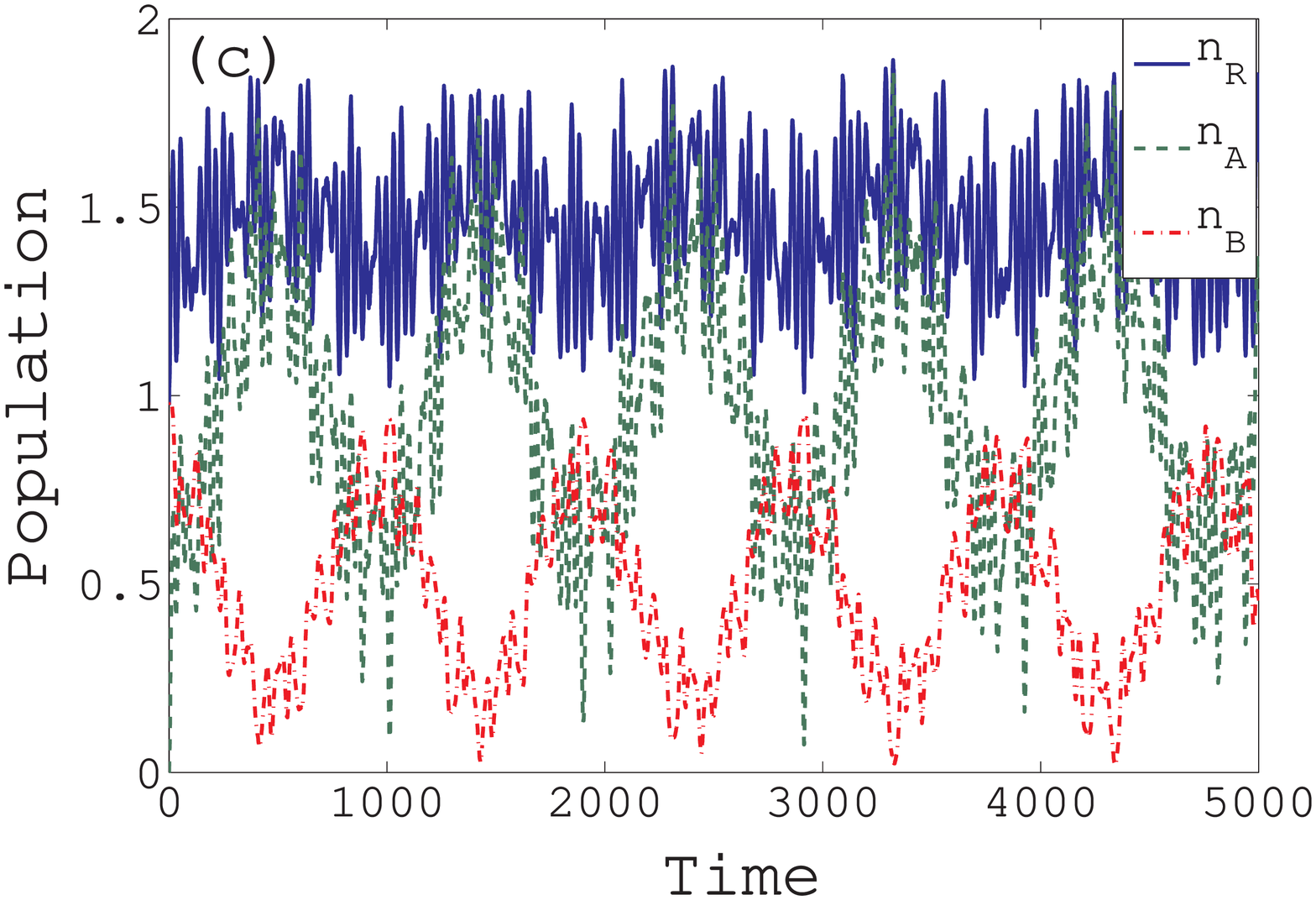}
\caption{(color online) Population of the right well $n_R$, $n_A$ and $n_B$ for $g_{AB}=25.0$, (a)  $g_A=0.0$, (b)  $g_A=5.0$, (c)  $g_A=20.0$  for the species-separated initial state. \label{cap:3p_gab25_sleft}}
\end{figure}

The avoided crossing present for the lowest lying states of the energy spectrum occurs also for higher values of $g_{AB}$ near $g_A \approx g_{AB}$. This results in a  similar dynamical behavior with respect to the dependence on $g_A$. Different is the case of high interactions $g_{AB}=25.0$, which is illustrated in Fig. \ref{cap:3p_gab25_sleft}. Here the tunneling period decreases substantially as $g_A$ takes larger values. The very smooth behavior for $g_A=0$ (Fig. \ref{cap:3p_gab25_sleft}(a)), where in principle only the lowest doublet $|AA,B\rangle \pm |B,AA\rangle$ contributes, changes to rapid small oscillations and erratic patterns as the intra-species interaction increases indicating that other higher lying states  are involved in the dynamics. The strong intra-species repulsion here serves again  as the principal destructor of the  correlated shuffling between the initial state and its mirror number-state. We can attribute the increase of the tunneling rate, to the increase of the density overlaps due to intra- and inter-species strong repulsion, in the line of the arguments provided in the discussion of Fig. \ref{cap:1p_density_left}.

As a last remark on the dynamics of the species separated initial state we would like to comment on the degree of correlation of the tunneling. Since the tunneling consists here in principle of a shuffling between $|AA,B\rangle$ and  $|B,AA\rangle$, the two species spent most of the time  in different wells. Therefore the probability to find B and A species in the same well remains always close to zero, while the A particles tunnel as a pair. Similar to  the previous section this behavior ceases to exist  in general for strong $g_A$ where single particle tunneling for the A species via excited states is induced. Note that for the so-far discussed cases of initial states, the destruction of the correlated tunneling behavior, (three bosons staying together; the two species  remaining separated), results from a strong increase of the intra-species interaction which drives the system beyond the simple number state dynamics ($|AAB,0\rangle$ $\Leftrightarrow $ $|0,AAB\rangle$ or $|AA,B\rangle$ $\Leftrightarrow$ $|B,AA\rangle$). We show next that such strong deviations from the initial state configuration can also be  achieved  for the situation of a  partially population imbalanced initial state but for a different reason.

\section{Partial population imbalanced initial state.}
\label{sub:3p_part}

A novel tunneling mechanism is encountered if the initial state is prepared such that the two wells share an equal  mean value of  the population of A atoms while the B atom is on the left well. This  initial state  we call \textit{partially population imbalanced} state. The behavior observed above namely the increase of the tunneling period with increasing $g_{AB}$ and its decrease with increasing $g_A$ can still be observed here. However, a major difference compared to the preceding cases arises in terms of the evolution of the different states which reflects itself in the corresponding time-evolution of the populations.

\begin{figure}

\includegraphics[width=0.45\columnwidth,keepaspectratio]{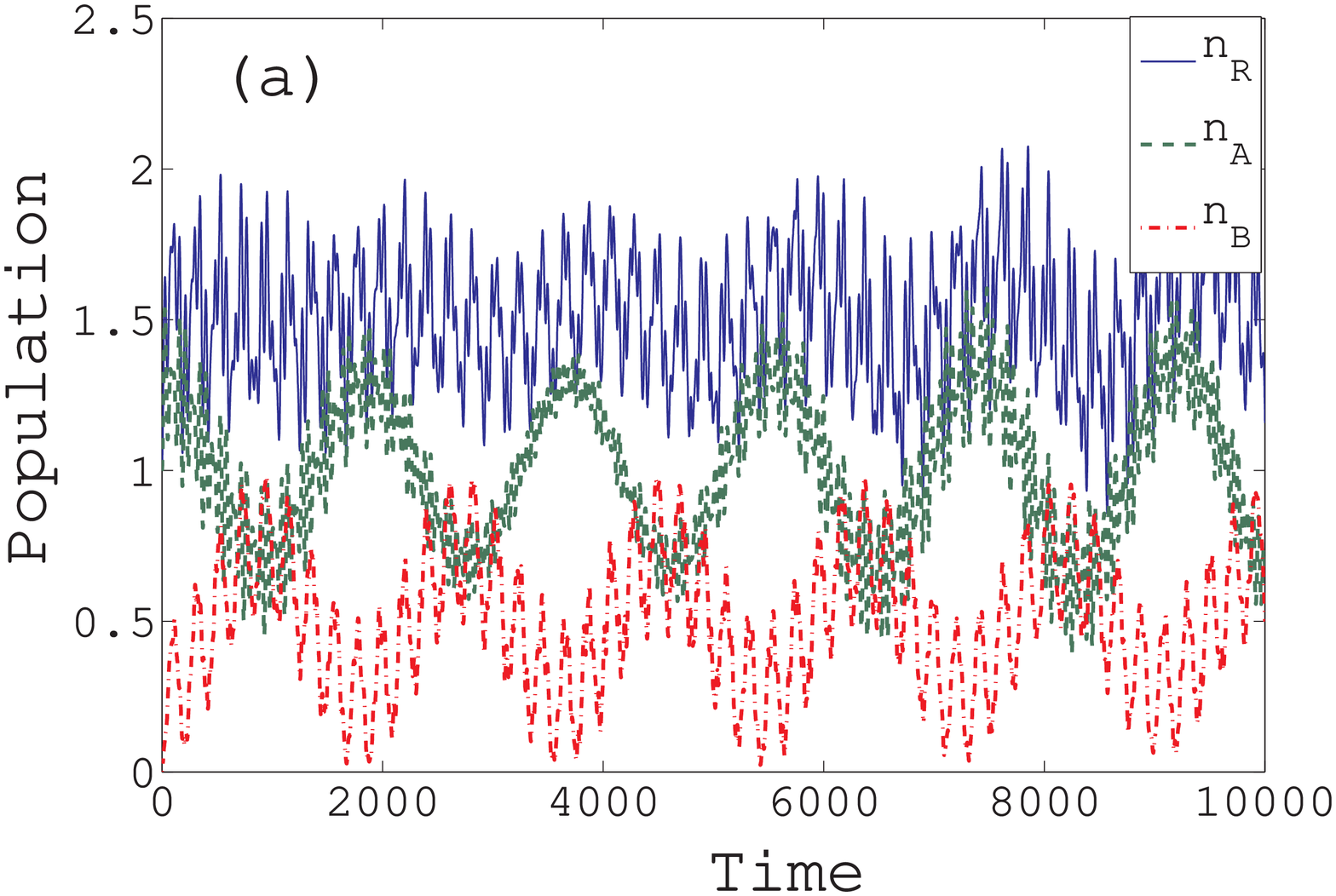}
\includegraphics[width=0.45\columnwidth,keepaspectratio]{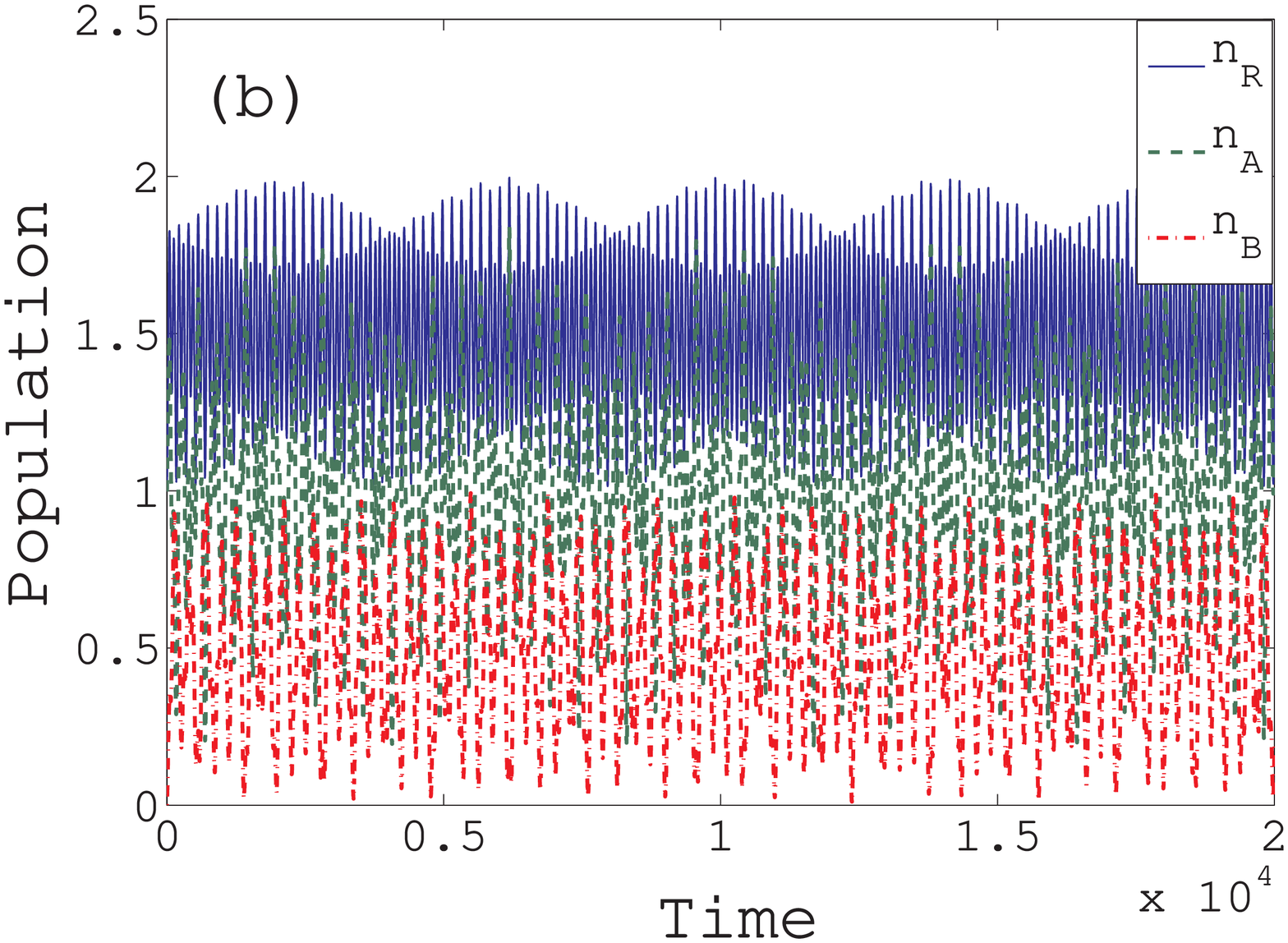}
\includegraphics[width=0.45\columnwidth,keepaspectratio]{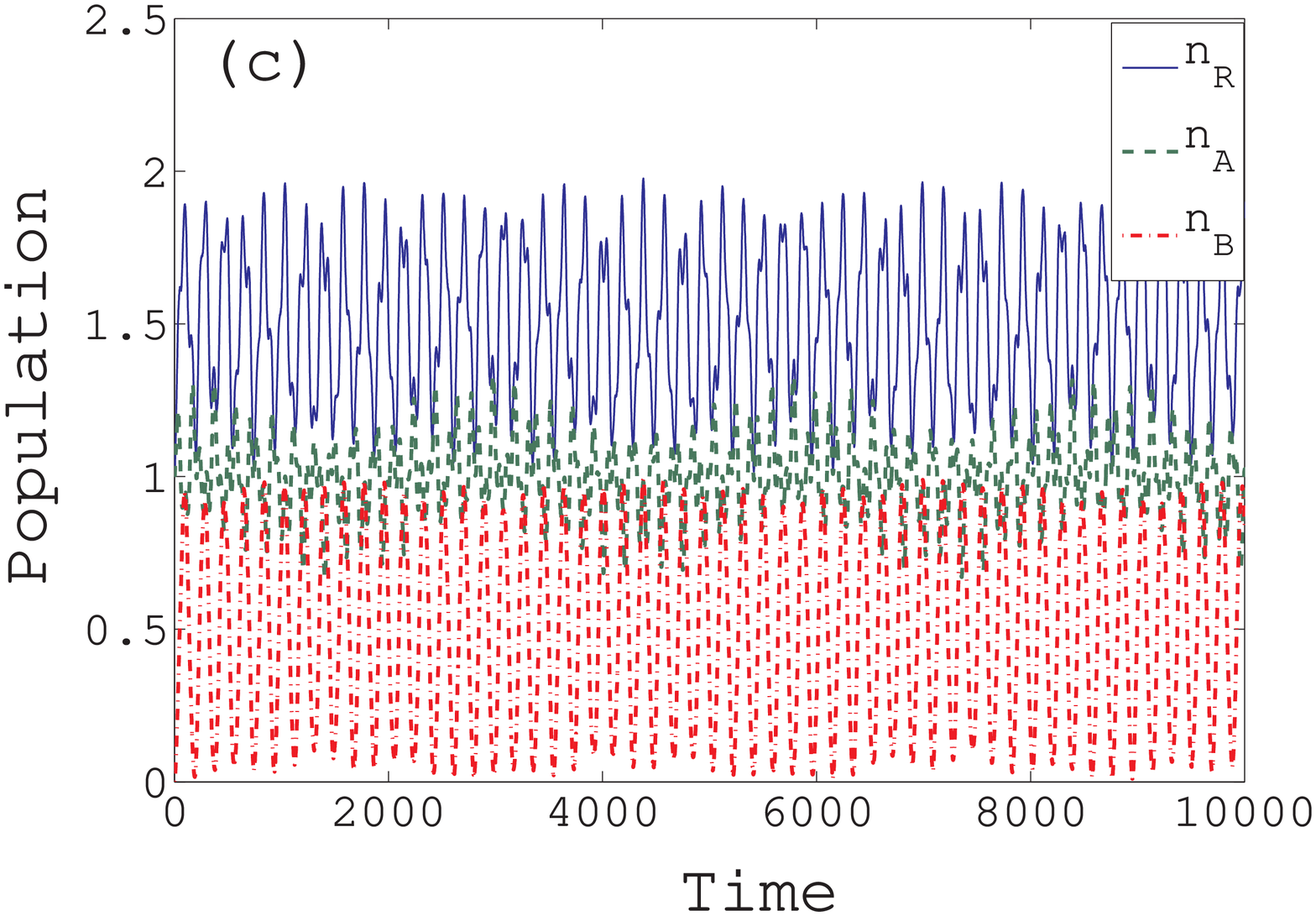}
\includegraphics[width=0.45\columnwidth,keepaspectratio]{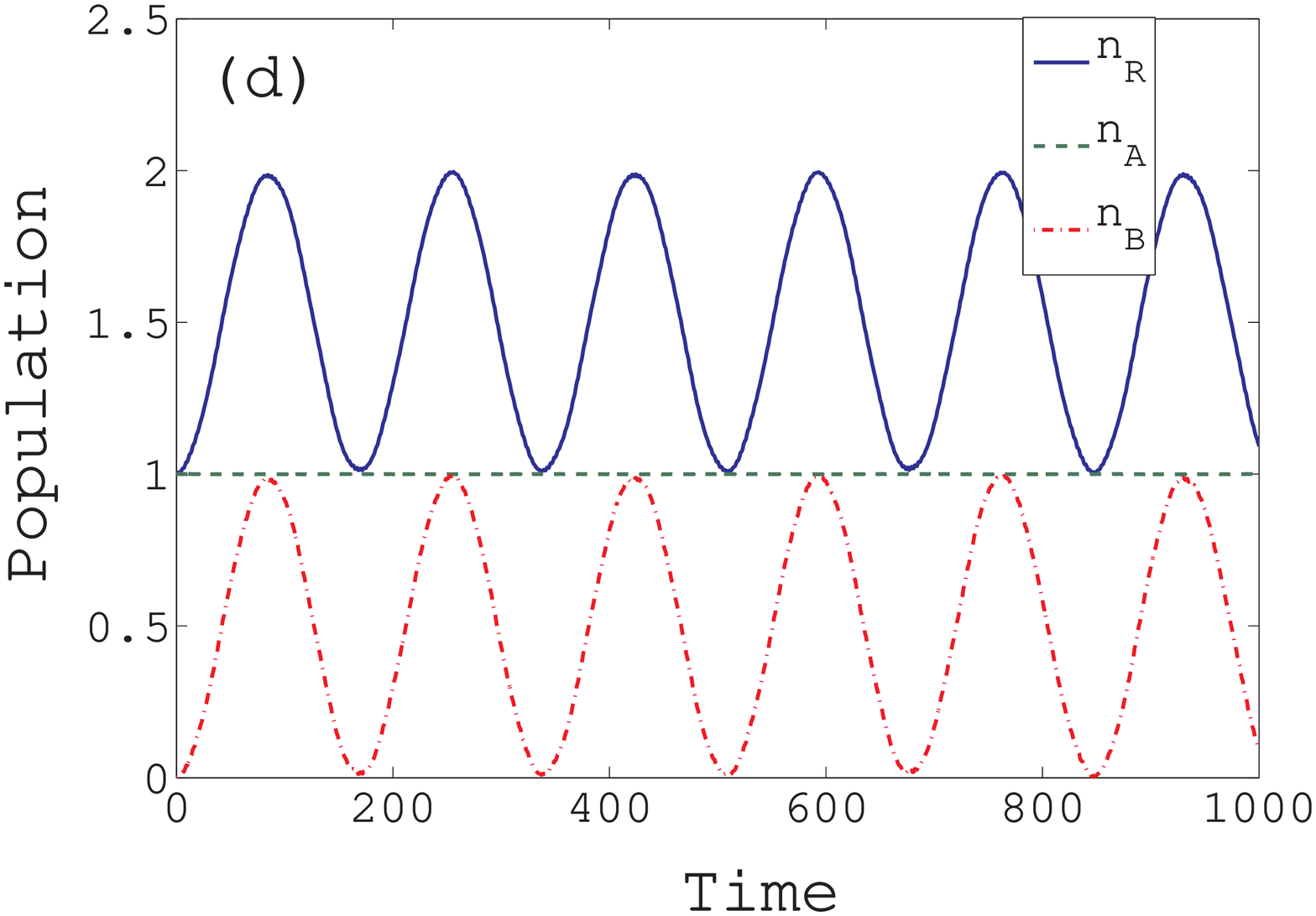}

\caption{(color online) Population in the right well $n_R$, $n_A$ and $n_B$ for $g_{AB}=0.2$ for (a) $g_A=0.0$, (b) $g_A=0.2$, (c) $g_A=0.3$ (d) $g_A=5.0$.\label{cap:3p_gab02_mleft}}
\end{figure}

In Fig. \ref{cap:3p_gab02_mleft} we show the populations $n_A$, $n_B$ and $n_R$ for $g_{AB}=0.2$. Naively, one would expect that the the B boson will undergo Rabi-oscillations on the background of the A bosons which should remain with equal population in each well. However this does not happen for $g_{AB}> g_A$. The envelope behavior of the A particle population i.e $n_A$ in Fig. \ref{cap:3p_gab02_mleft}(a) for $g_A=0$ first increases then decreases, indicating that the single A atom in the right well tunnels partially to the left well thus decreasing the population of the A particles in the right-well. The B boson on the other hand tunnels completely to the right well. This process is retained thereafter and is overall periodic. The envelope behavior is modulated by  high frequency oscillations of  significant amplitude involving a rapid tunneling between the two wells. As $g_A$ is increased from 0 to 0.2 the pattern becomes more irregular consisting mainly of a  constant envelope and shows rapid oscillations. The amplitude of the oscillation of $n_A$ remains large. When the intra-species interaction  strength $g_A=0.3$  becomes larger than the inter-species coupling $g_{AB}=0.2$ (Fig. \ref{cap:3p_gab02_mleft}(c)), the tunneling of A bosons is strongly suppressed. For even higher interactions $g_A=5.0$ (Fig. \ref{cap:3p_gab02_mleft}(d)), the A bosons are completely localized while the B boson undergoes Rabi oscillations between the two wells as one would  expect intuitively since the highly repulsive species A are initially in different wells.

\begin{figure}

\includegraphics[width=0.45\columnwidth,keepaspectratio]{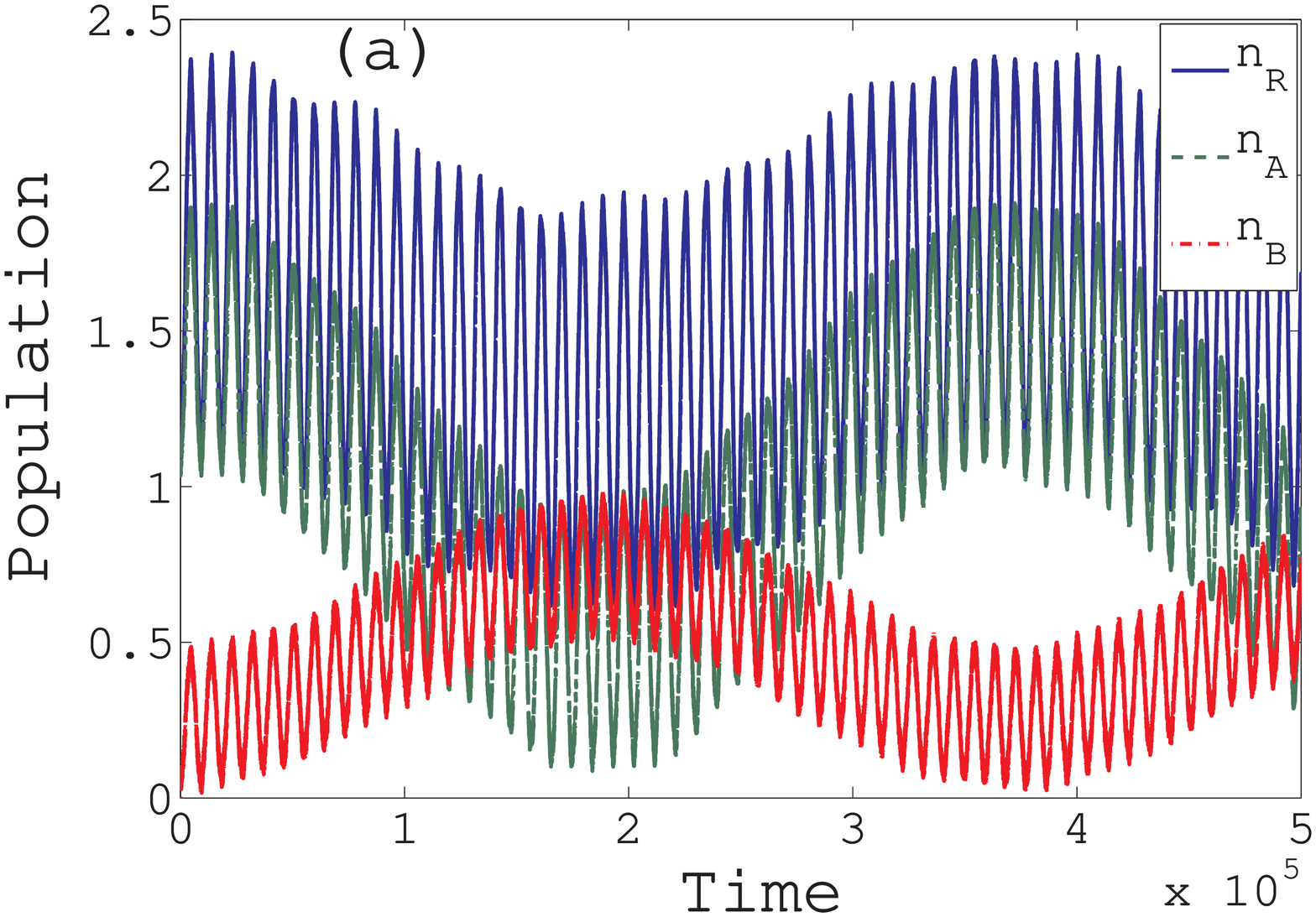}
\includegraphics[width=0.45\columnwidth,keepaspectratio]{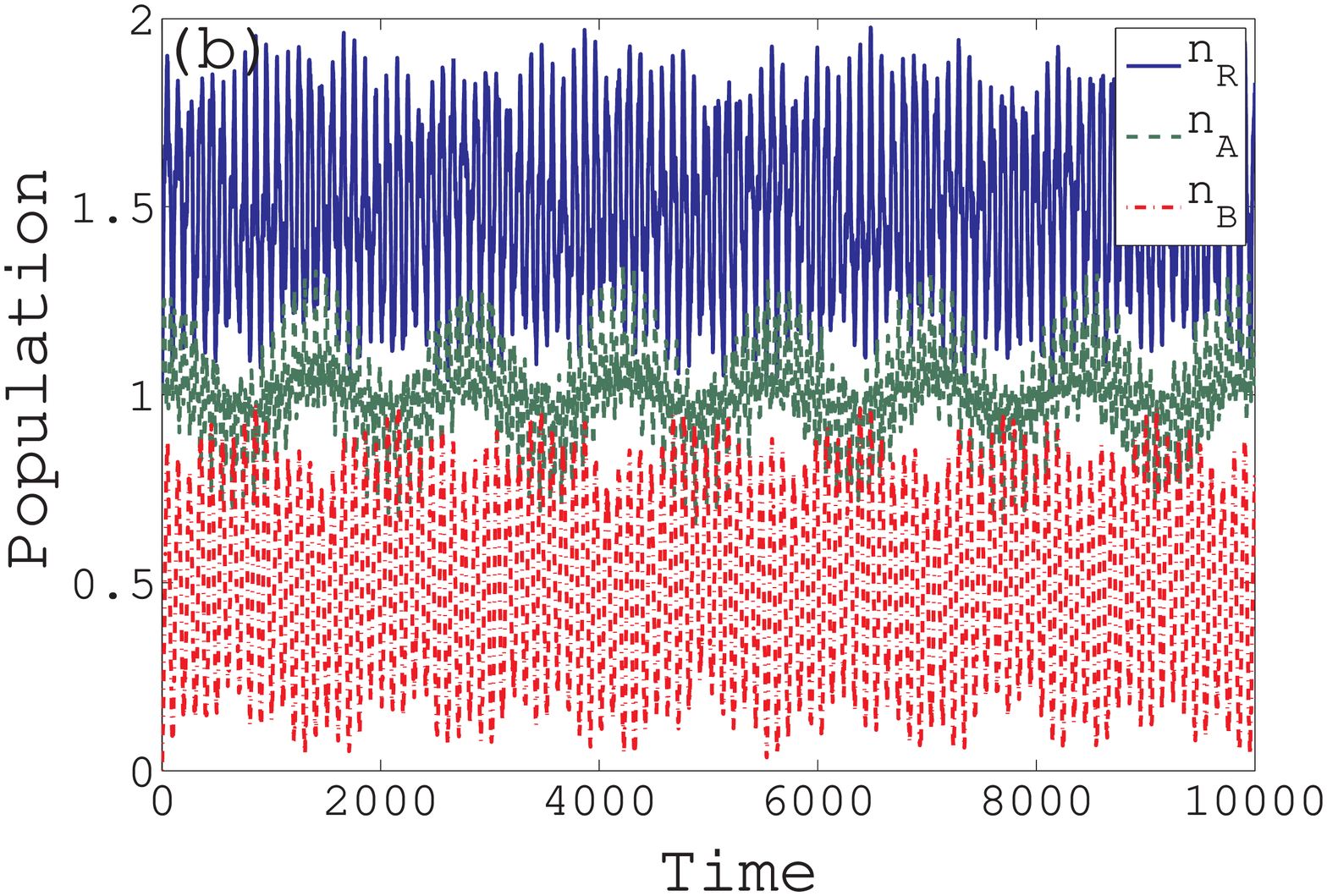}
\includegraphics[width=0.45\columnwidth,keepaspectratio]{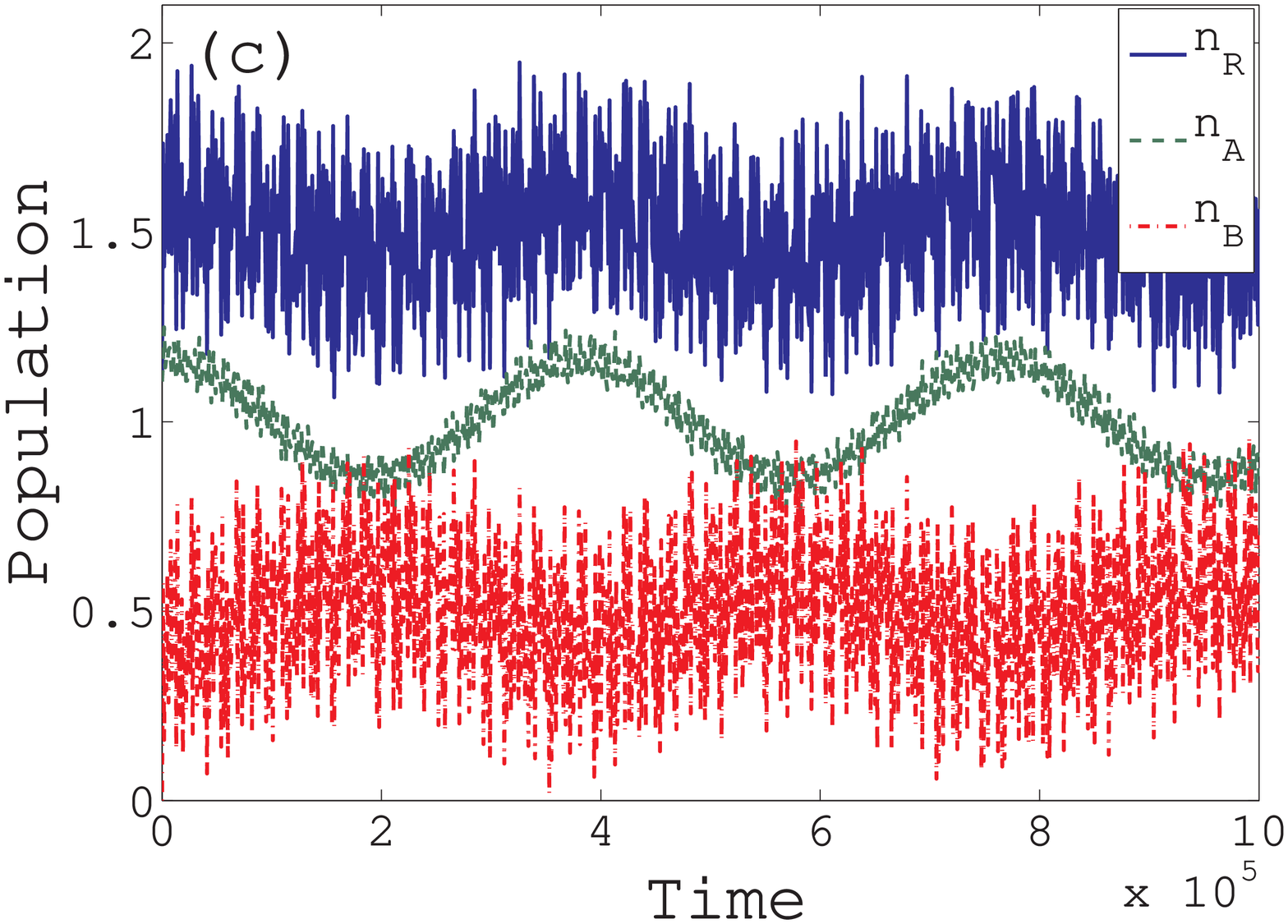}
\includegraphics[width=0.45\columnwidth,keepaspectratio]{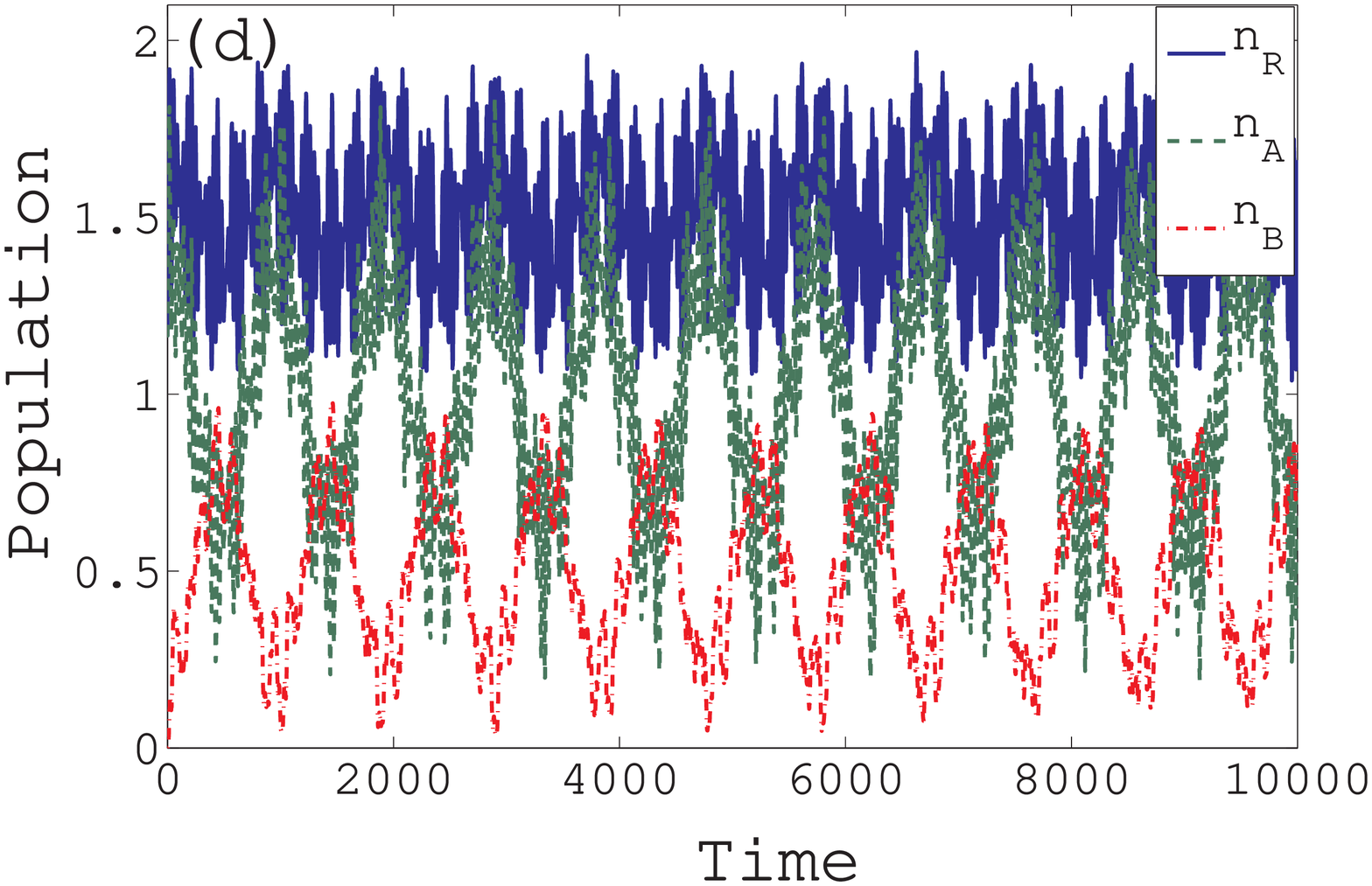}

\caption{(color online) Population in the right well  $n_R$, $n_A$ and $n_B$ for (a) $g_{AB}=5.0$ and $g_A=0.0$, (b) $g_{AB}=5.0$  and $g_A=4.0$, (c) $g_{AB}=25.0$ and $g_{A}=5.0$  (d) $g_{AB}=25.0$ and $g_{A}=20.0$ .\label{cap:3p_gab5_mleft}}
\end{figure}

The evolution of the dynamics shows further characteristics  for stronger interspecies interactions. Fig. \ref{cap:3p_gab5_mleft} presents the results for $g_{AB}=5.0$. For $g_A=0.0$  (Fig. \ref{cap:3p_gab5_mleft}(a)), there are two distinct oscillations for both $n_A$ and $n_B$: a fast fluctuation with significant amplitude for $n_B$ coupled to a large amplitude motion of $n_A$. Intuitively one can understand this behavior (seen also in the previous case) for large  $g_{AB}$ as follows: the  tunneling of the B boson to the right well pushes the A bosons to the left well due to the strong repulsion and vise versa leading to a counterflow type of dynamics. The fast oscillation of considerable amplitude for $n_A$ involves tunneling of a 'complete' A boson  and partial tunneling of a B boson between the wells. The origin of these oscillations can be understood via the number state decomposition of the initial state as will be explained below. Opposite to this, for $g_A=4.0$ (Fig. \ref{cap:3p_gab5_mleft}(b)), the tunneling of A bosons is considerably suppressed and the B boson undergoes a rapid oscillation between the wells. For even higher $g_A$ as before we get an almost complete suppression of the A boson tunneling while the B boson executes the same very fast oscillations.  

For very strong inter-species interaction $g_{AB}=25.0$, a similar pattern is seen for low $g_A$ (not shown) albeit with a much longer period. For quite strong $g_A=5.0$ there is a tendency for suppression of the tunneling of the A boson (Fig. \ref{cap:3p_gab5_mleft}(c)) which still oscillates but with a small amplitude.  Unlike  $g_{AB}=5.0$, increasing the interaction  to $g_A=20.0$ (Fig. \ref{cap:3p_gab5_mleft}(d)) does not reduce the tunneling of the A bosons but increases it approaching a 'fermionization' type behavior of the dynamics. 

To identify the underlying dynamical mechanisms leading to the above observations, we first note that the initial state in this case is not necessarily a pure number-state $|AB,A\rangle$ but is a linear combination of the number states: $|AAB,0\rangle$,  $|AB,A\rangle$ and  $|B,AA\rangle$  maintaining the required population balance of the initial state (equal population of A bosons in each well and the B boson in the left). 

For this initial setup the tunneling dynamics consists of transferring the atoms between the initial state  and a target state composed of the number-states $|0,AAB\rangle$, $|A,AB\rangle$ and  $|AA,B\rangle$.  For $g_{AB} \gg g_A$, the number-state $|AA,B\rangle$ represents the  dominant contribution to the target state and thus the dynamics consists of transferring the atoms between the initial state and the configuration $|AA,B\rangle$. As a result we have a  tunneling of the B boson to the right well and of a single A boson to the left well which we can observe in  the envelope behavior of $n_A$ and $n_B$ of Fig.\ref{cap:3p_gab02_mleft}(a) and more prominently in Fig. \ref{cap:3p_gab5_mleft}(a).  The faster oscillations are  the result of  the contributions from the states  $|0,AAB\rangle$ and $|A,AB\rangle$. For  $g_{AB} \approx g_A$, we have contributions of approximately the same magnitude  from almost all the number states leading to  Josephson like oscillations. However, for $g_{AB} \ll g_A$, the dominant contribution of the target state is $|A,AB\rangle$ and thus the system shows a transfer between the initial state and the state $|A,AB\rangle$. Therefore  the A bosons are effectively localized  while the B bosons undergo Rabi oscillations between the wells.

\section{Conclusion and Outlook}

We have investigated the tunneling dynamics of a strongly correlated few body bosonic binary mixture in a one-dimensional double-well covering the complete range of intra- and inter-species interaction. Our focus  is the interplay of the inter- and intra-species correlations and  their impact on the dynamics. We observe that  the tunneling period increases drastically as the inter-species interaction $g_{AB}$ increases which is due to  quasi-degenerate symmetric states contributing primarily to the dynamics. This effect is quite general and observed here for different initial configurations.  

The intra-species coupling $g_A$ possesses a  different impact on the behavior of the dynamics, depending on the strength  $g_{AB}$ as well as on the initial state. The general trend is that for large $g_A$ the overlap of localized wave functions of contributing states becomes larger and thus the effective tunneling coupling is increased leading to higher tunneling frequencies.  For low interactions though different behavior is encountered for different setups. For a completely imbalanced initial state, for instance, we observe that for small values of $g_{AB}$, the tunneling period increases  as we increase $g_A$ in the weak interaction regime. However for larger values of $g_A$, the tunneling period reduces with increasing $g_A$. This behavior is not seen for the species-separated initial condition. In the latter case, we observe a  minimal period at $g_A=g_{AB}$ which is a manifestation of an avoided crossing in the spectrum. 

Concerning the different initial states of the ensemble the complete population imbalanced state exhibits generically a completely correlated tunneling process for the A and B species which breaks only for large values of $g_A$ leading to an attempted single particle tunneling and independent fermion-like behavior. For the species separated scenario the two species tend to stay in opposite wells when the inter-species repulsion is large, a behavior which alters only if $g_A$ becomes also large. For the partially population imbalanced case where the mean population of A atoms in each well $n_A = 1$, although one would intuitively expect that the A particles remain in different wells due to their initial preparation, this happens only if the interaction between them is considerably large. In the other cases the A particles undergo oscillations and the initially mixed state where an A and a B boson coexist in the same well can turn into a separated state for which the A and B species reside in different wells. 

Understanding the fundamental effects and mechanisms of the tunneling dynamics in strongly correlated bosonic mixtures on a few body level can be seen as a starting point to realize systems such as bosonic transistors or to create schemes for selective transport of individual bosonic component in  reservoir-sink systems as well as for studies of entanglement and statistical properties of mixed ensembles. Further considerations could include higher number of particles or species and effects of  parameters which differentiate the two species.   

\appendix
\section{Computational Method MCTDH \label{sec:mctdh}}

Our goal is to study the bosonic quantum dynamics for weak to strong interactions in a numerically exact fashion. This is computationally challenging and can be achieved only for few atom system.
Our approach is the Multi-Configuration Time Dependent Hartree (MCTDH) method \cite{meyer90,beck00} being a wave packet dynamical tool known for its outstanding efficiency in high dimensional applications.

The principle idea is to solve the time-dependent Schr\"odinger equation 
\\
\begin{center}
$i\dot{\Psi}(t) = H\Psi(t)$ \\
\end{center}
as an initial value problem by expanding
the solution in terms of Hartree products $\Phi_J \equiv {\varphi_j}_1 \otimes \ldots \otimes {\varphi_j}_N$ :

\begin{equation}\Psi(t) = \sum_J A_J(t)\Phi_J(t).\label{eq:mctdh}\end{equation}

The unknown single particle functions $\varphi_j(j=1,...,n$, where $n$ refers to the total number of single particle functions used in the calculation) are in turn  represented in a fixed primitive basis implemented on a grid. The correct bosonic permutation symmetry is obtained by symmetrization of the expansion coefficient $A_J$.
Note that in the above expansion, not only are the coefficients $A_J$ time dependent but  also the single particle functions $\varphi_j$.
Using the  Dirac-Frenkel variational principle, one can  derive the equations of motion for both $A_J$ and $\Phi_J$.
Integrating these differential equations of motion  gives us  the time evolution of the system via (\ref{eq:mctdh}).
This has the advantage that the basis $\Phi_J(t)$ is variationally optimal at each time $t$. Thus it can be kept relatively small, rendering the procedure more efficient.

Although MCTDH is designed primarily for time dependent problems, it is also possible to compute  stationary states. For this purpose the \textit{relaxation} method is used \cite{kos86:223}. The key idea is to propagate a wave function $\Psi_0$ by the non-unitary operator $e^{-H\tau}$. As $\tau \rightarrow \infty$, this exponentially damps out any contribution but that stemming from  the true ground state like $e^{-(E_m - E_0)\tau}$. In practice one relies upon a more sophisticated scheme called the \textit{improved relaxation} \cite{mey03:251,meyer06} which is much more robust especially for excited states.  Here $\langle\Psi\vert H \vert \Psi \rangle$ is minimized with respect to both the coefficients $A_J$ and the orbitals $\varphi_j$. The effective eigenvalue problems thus obtained are then solved iteratively by first solving $A_J$ with fixed orbital $\varphi_j$ and then optimizing $\varphi_j$ by propagating them in imaginary time over a short period. This cycle is then repeated.

We note here that  the computation of  very long tunneling times using the MCTDH propagation scheme is numerically impractical. For these cases we computed the dynamics through the expansion of few-body eigenstates. Moreover, for extremely close quasi-degenerate states, convergence is difficult. In these cases a simultaneous relaxation of a whole set of these eigenstates keeping them orthogonal is performed by a method known as  block relaxation.

\acknowledgments{ B.C. gratefully acknowledges the financial and academic support from the  International Max-Planck Research School for Quantum Dynamics in Physics, Chemistry and Biology. L.C. gratefully thanks the Alexander von Humboldt Foundation (Germany) for a fellowship. P.S. acknowledges financial support by the Deutsche Forschungsgemeinschaft (DFG).}


\begin{thebibliography}{10}

\bibitem{pitaevskii}
L. Pitaevskii and S. Stringari, { Bose-Einstein Condensation} (Oxford
  University Press, Oxford, 2003).

\bibitem{pethick}
C.~J. Pethick and H. Smith, { Bose-Einstein condensation in dilute gases}
  (Cambridge University Press, Cambridge, 2008).

\bibitem{bloch07}
I. Bloch, J. Dalibard, and W. Zwerger, Rev. Mod. Phys. {\bf 80},  885  (2008).

\bibitem{buluta09}
I. Buluta and F. Nori, Science
  {\bf 326},  108  (2009).

\bibitem{greiner02}
M. Greiner {\it et~al.}, Nature
  {\bf 415},  39  (2002).

\bibitem{lewenstein07}
M. Lewenstein, A. Sanpera, D. Bogdan, A. Sen, and U. Sen, Adv. Phys. {\bf 56},
  243  (2007).

\bibitem{kierig08}
E. Kierig {\it et~al.}, Phys. Rev. Lett. {\bf 100},  190405  (2008).

\bibitem{chin10}
C. Chin,  and R. Grimm, P. Julienne, and E. Tiesinga, 
Rev. Mod. Phys. {\bf 82},  1225  (2010).


\bibitem{Olshanii1998a}
M. Olshanii, Phys. Rev. Lett. {\bf 81},  938  (1998).

\bibitem{myatt97}
C.~J. Myatt {\it et~al.}, Phys. Rev. Lett. {\bf 78},  586  (1997).

\bibitem{hall98}
D.~S. Hall {\it et~al.}, Phys. Rev. Lett. {\bf 81},  1539  (1998).

\bibitem{maddaloni00}
P. Maddaloni {\it et~al.}, Phys. Rev. Lett. {\bf 85},  2413  (2000).

\bibitem{modugno02}
G. Modugno {\it et~al.}, Phys. Rev. Lett. {\bf 89},  190404  (2002).

\bibitem{catani08}
J. Catani {\it et~al.}, Phys. Rev. A {\bf 77},  011603  (2008).

\bibitem{cazalilla03}
M.~A. Cazalilla and A.~F. Ho, Phys. Rev. Lett. {\bf 91},  150403  (2003).

\bibitem{alon06}
O.~E. Alon, A.~I. Streltsov, and L.~S. Cederbaum, 
Phys. Rev. Lett. {\bf 97}, 230403  (2006).

\bibitem{mishra07}
T. Mishra, R.~V. Pai, and B.~P. Das, Phys. Rev. A {\bf 76},  013604  (2007).

\bibitem{roscilde07}
T. Roscilde and J.~I. Cirac, Phys. Rev. Lett. {\bf 98},  190402  (2007).

\bibitem{kleine08}
A. Kleine, C. Kollath, I.~P. McCulloch, T. Giamarchi, and U. Schollw\"{o}ck,
  Phys. Rev. A {\bf 77},  013607  (2008).

\bibitem{girardeau07}
M.~D. Girardeau and A. Minguzzi, Phys. Rev. Lett. {\bf 99},  230402  (2007).

\bibitem{zoellner08b}
S. Z{\"o}llner, H.-D. Meyer, and P. Schmelcher, Phys. Rev. A {\bf 78},  013629
  (2008).

\bibitem{hao08}
Y. Hao and S. Chen, Eur. Phys. J. D {\bf 51},  261  (2009).

\bibitem{hao09}
Y. Hao, Y. Zhang, X.-W. Guan, and S. Chen, Phys. Rev. A {\bf 79},  033607
  (2009).

\bibitem{tempfli09}
E. Tempfli, S. Z{\"o}llner, and P. Schmelcher, New J. Phys. {\bf 11},  073015
  (2009).

\bibitem{mathey09}
A. Hu, L. Mathey, I. Danshita, E. Tiesinga, C. J. Williams, C. W. Clark, Phys. Rev. A {\bf 80},  023619
  (2009).


\bibitem{kinoshita04}
T. Kinoshita, T. Wenger, and D.~S. Weiss, Science {\bf 305},  1125  (2004).

\bibitem{paredes04}
B. Paredes {\it et~al.}, Nature {\bf 429},  277  (2004).


\bibitem{girardeau60}
M. Girardeau, J. Math. Phys. {\bf 1},  516  (1960).

\bibitem{salgueiro06}
A.~N. Salgueiro {\it et~al.}, Eur. Phys. J. D {\bf 44},  537  (2007).


\bibitem{dounasfrazer07b}
D.~R. Dounas-Frazer, A.~M. Hermundstad, and L.~D. Carr, Phys. Rev. Lett. {\bf
  99},  200402  (2007).


\bibitem{wang08}
L. Wang, Y. Hao, and S. Chen, Eur. Phys. J. D {\bf 48},  229  (2008)

\bibitem{albiez05}
M. Albiez, R. Gati, J. F\"olling, S. Hunsmann, M. Cristiani, and
M. K. Oberthaler, Phys. Rev. Lett. {\bf 95},  010402  (2005).

\bibitem{anker05}
T. Anker, M. Albiez, R. Gati, S. Hunssmann, B. Eiermann,
A. Trombettoni, and M. K. Oberthaler, Phys. Rev. Lett. {\bf 94},  020403  (2005).

\bibitem{kuang00}
L-M. Kuang and Z-W Ouyang. Phys. Rev. A \textbf{% 
61}, 023604 (2000).

\bibitem{xu08}  
X. Q. Xu, L. H. Lu, Y. Q. Li, Phys. Rev. A \textbf{%
78 }, 043609 (2008).

\bibitem{mazzarella09}  
G. Mazzarella, M. Moratti, L. Salasnich, M. Salerno, F.
Toigo, J. Phys. B: At. Mol. Opt.\textbf{42},
125301 (2009).

\bibitem{satija09}  
I. I. Satija, R. Balakrishnan, P. Naudus, J. Heward, M.
Edwards, C. W. Clark, Phys. Rev. A \textbf{79},
033616 (2009).

\bibitem{diaz09} 
 B. Julia-Diaz, M. Guilleumas, M. Lewenstein, A. Polls, A.
Sanpera, Phys. Rev. A \textbf{80 }, 023616 (2009).

\bibitem{sun09}  
B. Sun, M. S. Pindzola, Phys. Rev. A \textbf{80 },
 033616 (2009).

\bibitem{naddeo10}
A Naddeo, R Citro, J. Phys. B: At. Mol. Opt. Phys {\bf 43},  135302 (2010).

\bibitem{mathey11}
A. Hu, L. Mathey, E. Tiesinga, I. Danshita,  C. J. Williams, C. W. Clark,
arXiv:1103.3513v2 (2011)


\bibitem{pflanzer09}
A.C. Pflanzer, S. Z{\"o}llner and P. Schmelcher, J. Phys. B  (FT) {\bf 42}, 231002 (2009).

\bibitem{pflanzer10}
A.C. Pflanzer, S. Z{\"o}llner and P. Schmelcher, Phys. Rev. A {\bf 81},  023612
  (2010).

\bibitem{widera11}
A. Widera, W. Alt and D. Meschede,	J. Phys.: Conf. Ser. \textbf{264}, 012021 (2011)

\bibitem{zoellner07a}
S. Z{\"o}llner, H.-D. Meyer, and P. Schmelcher, Phys. Rev. Lett. {\bf 100},
  040401  (2008)

\bibitem{zoellner08}
S. Z{\"o}llner, H.-D. Meyer, and P. Schmelcher, Phys. Rev. A {\bf 78},  013621
  (2008).


\bibitem{meyer90}
H.-D. Meyer, U. Manthe, and L.~S. Cederbaum, Chem.\ Phys.\ Lett. {\bf 165},  73
   (1990).

\bibitem{beck00}
M.~H. Beck, A. J{\"a}ckle, G.~A. Worth, and H.-D. Meyer, Phys.\ Rep. {\bf 324},
   1  (2000).

\bibitem{kos86:223}
R. Kosloff and H. Tal-Ezer, Chem.\ Phys.\ Lett. {\bf 127},  223  (1986).

\bibitem{mey03:251}
H.-D. Meyer and G.~A. Worth, Theor.\ Chem.\ Acc. {\bf 109},  251  (2003).

\bibitem{meyer06}
H.-D. Meyer, F.~L. Qu\'{e}r\'{e}, C. L\'{e}onard, and F. Gatti, Chem. Phys.
  {\bf 329},  179  (2006).


\end{thebibliography}
\end{document}